\documentclass[preprint,12pt,authoryear]{elsarticle}

\usepackage{amssymb}
\usepackage{amsmath}

\journal{Environmental International}

\begin{document}

\begin{frontmatter}



\title{Integrating Temporal Disaggregation and Distributed Lag Nonlinear Models for Bayesian Spatio-Temporal Disease Mapping with High-Resolution Environmental Exposures} 


\author[inst1, inst2]{Alejandro Rozo Posada} 
\author[inst1]{Maxime Fajgenblat}
\author[inst2]{Christel Faes}
\author[inst3]{James Colborn} 
\author[inst4]{Emanuele Giorgi}
\author[inst3]{Baltazar Candrinho}
\author[inst1,inst2]{Thomas Neyens}

\affiliation[inst1]{organization={Leuven Biostatistics and Statistical Bioinformatics Centre (L-Biostat),
Department of Public Health and Primary Care, KU Leuven},
            addressline={Kapucijnenvoer 7}, 
            city={Leuven},
            postcode={3000}, 
            state={Flemish Brabant},
            country={Belgium}}

\affiliation[inst2]{organization={Data Science Institute, Hasselt University},
addressline={Agoralaan - gebouw D}, 
city={Diepenbeek},
postcode={3590}, 
state={Limburg},
country={Belgium}}

\affiliation[inst3]{organization={Clinton Health Access Initiative},
city={Maputo},
country={Mozambique}}

\affiliation[inst4]{organization={National Malaria Control Program, Ministry of Health},
city={Maputo},
country={Mozambique}}

\affiliation[inst4]{organization={Department of Applied Health Sciences, University of Birmingham},
addressline={Edgbaston}, 
city={Birmingham},
postcode={B15 2TT}, 
state={West Midlands},
country={United Kingdom}}

\begin{abstract}
Environmental conditions are major drivers of malaria transmission, however epidemiological analyses are often constrained by temporal misalignment between health outcomes reported at coarse time scales and environmental exposures available at much finer temporal resolutions. Conventional approaches typically aggregate high-resolution environmental data to match the temporal resolution of health outcomes, potentially obscuring delayed and nonlinear exposure-response relationships. This paper proposes a Bayesian spatio-temporal modeling framework that addresses this limitation by introducing a latent daily disease process linked to observed monthly malaria counts through temporal disaggregation. The framework jointly incorporates distributed lag nonlinear models to characterize delayed and nonlinear climatic effects, spatio-temporal structured random effects to account for residual dependence, and district-level intervention covariates within an unified hierarchical model. \\

The methodology was applied to monthly malaria surveillance data from 161 districts in Mozambique between 2017 and 2024, together with daily temperature, precipitation, and relative humidity, monthly vegetation index, elevation, and malaria intervention data. Predictive performance was evaluated against a conventional monthly model using the same set of covariates. The proposed framework achieved improved predictive accuracy and uncertainty quantification while exploiting the full temporal resolution of the environmental information. Estimated exposure-response relationships identified nonlinear associations between climatic variability and malaria incidence that were consistent with current epidemiological understanding, including an optimal temperature range for transmission, increasing risk with positive vegetation anomalies, and nonlinear precipitation effects. \\

By avoiding temporal aggregation of environmental exposures, the proposed framework provides a flexible and coherent approach for investigating delayed environmental effects using routinely collected surveillance data. Although illustrated using malaria in Mozambique, the methodology is readily applicable to other environmentally sensitive diseases where health outcomes and environmental exposures are observed at different temporal resolutions.
\end{abstract}




\end{frontmatter}



\section{Introduction}
\label{Intro}
Malaria remains one of the major global public health challenge. According to the  \cite{WHO2025Malaria}, an estimated 282 million cases and 597,000 deaths occurred worldwide in 2024, with the African Region accounting for over 90\% of the global burden. Although substantial progress has been achieved through vector control, improved diagnosis, and effective treatment, recent reductions in malaria incidence have stopped. Since 2015 malaria case incidence has increased by 8.5\%, contrary to the reduction observed between 2000 and 2015 of 25\%. These figures highlight among others the need for even stronger surveillance systems capable of supporting timely and evidence-based public health interventions. Accurate surveillance is particularly important because malaria transmission varies considerably across space and time in response to environmental conditions, population characteristics, and control measures. \\

Among the factors influencing malaria transmission, environmental conditions are particularly valuable because they can be monitored continuously over large geographical areas using meteorological observations and satellite products. These conditions play a fundamental role in malaria transmission because they directly influence both mosquito ecology and parasite development \citep{alcayna2025identifying}. Specifically, temperature regulates mosquito survival, biting activity, and the rate of parasite development within the vector \citep{yamba2023climate}. Precipitation determines the availability of aquatic breeding habitats, although excessive rainfall may reduce vector populations by destroying larval habitats. Relative humidity affects mosquito survival by reducing desiccation \citep{santos2022neglected}, while vegetation and topographic characteristics influence habitat suitability and the spatial distribution of vector populations \citep{ekpa2023spatio}. These environmental variables therefore provide an important source of information for anticipating changes in malaria risk. As environmental datasets have become increasingly available at fine spatial and temporal resolutions, the challenge is no longer obtaining environmental information, but developing statistical methods that make the most effective use of it for disease surveillance and prediction. \\ 

Despite the increasing availability of environmental data at fine temporal resolutions, an important challenge remains because environmental and epidemiological data are typically collected at different temporal scales. Environmental conditions can now be monitored daily, or even hourly, through meteorological observations and remote sensing products, whereas routine malaria surveillance commonly reports aggregated case counts at weekly or monthly intervals. To reconcile these different temporal scales, most studies aggregate environmental variables to match the temporal resolution of the health outcome before analysis. This practice is widespread in malaria epidemiology, including studies from Kenya \citep{beloconi2023malaria}, Mozambique \citep{armando2023climate}, and broader analyses across sub-Saharan Africa \citep{yamba2023climate}. While this approach enables analyses using routinely collected surveillance data, it averages environmental conditions that may have markedly different biological consequences for malaria transmission. As a result, short-term variability and extreme events can be obscured, while delayed exposure-response relationships become more difficult to estimate accurately, limiting the extent to which available environmental information can be used for inference and prediction \citep{economou2026modelling}. \\ 

The loss of information associated with temporal aggregation has motivated a range of methodological developments designed to better exploit environmental data and characterize climate–health relationships across multiple temporal scales. One line of research has focused on understanding how exposure–response relationships vary across different temporal scales. For example, \cite{masselot2022data} proposed several methodological strategies to separate environmental signals operating at different temporal frequencies and evaluate their individual contributions to health outcomes, however, these approaches do not explicitly address the temporal mismatch between environmental exposures and health outcomes, nor do they reconstruct the underlying fine-scale health process from aggregated observations. \\

More recently, methodological developments have begun to explicitly address the temporal mismatch between high-frequency environmental exposures and lower-frequency health outcomes. For instance, \cite{economou2026modelling} proposed a likelihood-based framework that exploits the infinite divisibility of the Poisson distribution to derive the likelihood of aggregated health counts from an underlying fine-resolution Poisson process. The framework further incorporates distributed lag non-linear models (DLNMs) \citep{gasparrini2010dlnm} to characterize potentially non-linear and delayed associations between climatic exposures and health outcomes. These relationships are represented using flexible penalized spline functions, allowing the complete sequence of high-frequency environmental exposures to be retained rather than aggregated to the temporal resolution of the health outcome. \\

An alternative approach was recently proposed by \cite{shukla2026mixed} through the mixed-frequency distributed lag nonlinear model (mf-DLNM), which extends the distributed lag nonlinear model (DLNM) framework \citep{gasparrini2010dlnm} to settings where environmental exposures and health outcomes are observed at different temporal resolutions. Rather than reformulating the likelihood, the mf-DLNM modifies the construction of the cross-basis so that the complete sequence of high-frequency environmental observations and their lagged contributions are directly linked to lower-frequency health outcomes. This enables nonlinear and delayed exposure–response relationships to be estimated while preserving the temporal variability of the environmental data. However, the methodology has so far been evaluated primarily through simulation studies and remains to be validated in complex infectious disease applications. \\

Addressing the temporal mismatch between environmental exposures and health outcomes requires analytical frameworks capable of integrating information collected at different temporal resolutions. However, environmental epidemiology, and infectious disease surveillance in particular, presents additional challenges, including delayed and nonlinear exposure–response relationships, spatio-temporal dependence, and multiple sources of uncertainty that should ideally be modeled jointly. Existing methodological developments have largely addressed these challenges separately. While recent approaches have proposed solutions for temporal misalignment or mixed-frequency exposure modeling \citep{economou2026modelling,shukla2026mixed}, they do not simultaneously accommodate the spatial dependence, temporal dynamics, and hierarchical uncertainty structure characteristic of routine surveillance data. Likewise, Bayesian spatio-temporal disease mapping models effectively capture spatial and temporal dependence but generally rely on temporally aggregated environmental covariates. Consequently, there remains a need for integrated statistical frameworks that preserve the temporal information contained in environmental exposures while coherently modeling the multiple processes underlying disease transmission. \\

This study proposes a unified Bayesian spatio-temporal framework that addresses these challenges simultaneously. Building on the infinite divisibility property of count distributions, whereby sums of independent random variables remain within the same distributional family under appropriate parameterization, the framework formulates observed monthly malaria counts as aggregations of an underlying latent daily disease process. This formulation links the environmental and epidemiological data within a coherent hierarchical model without requiring temporal aggregation of the environmental exposures. The probabilistic formulation naturally propagates uncertainty associated with the unobserved daily disease process while allowing the complete sequence of daily environmental observations to contribute to inference. Within this framework, distributed lag nonlinear models characterize delayed and nonlinear exposure–response relationships, while spatial random effects account for residual spatial dependence and additional model components accommodate seasonal variation and malaria intervention measures. The resulting framework provides an integrated approach for environmental epidemiology that jointly addresses temporal misalignment, nonlinear lagged environmental effects, spatial dependence, and uncertainty propagation. \\

The proposed framework is demonstrated using routinely collected malaria surveillance data from Mozambique, where monthly district-level malaria counts are linked with daily environmental observations. This application provides a realistic setting for evaluating whether explicitly accounting for the temporal mismatch between environmental exposures and disease surveillance improves predictive performance and yields more informative estimates of climate-malaria relationships. More broadly, the proposed framework offers a general approach for analyzing environmentally driven diseases whenever exposure and outcome data are collected at different temporal resolutions. \\

The remainder of this paper is organized as follows. Section~\ref{methods} introduces the malaria surveillance data and covariates used in the analysis and describes the proposed modeling framework. Section~\ref{ch4_results} presents the results, while Section~\ref{ch4_discussion} discusses the main findings and their implications. Finally, Section~\ref{ch4_conclu} provides the concluding remarks.

\section{Material and Methods}
\label{methods}
\subsection{Malaria incidence data}

\begin{figure}
  \centering
   \includegraphics[width=0.3\linewidth]{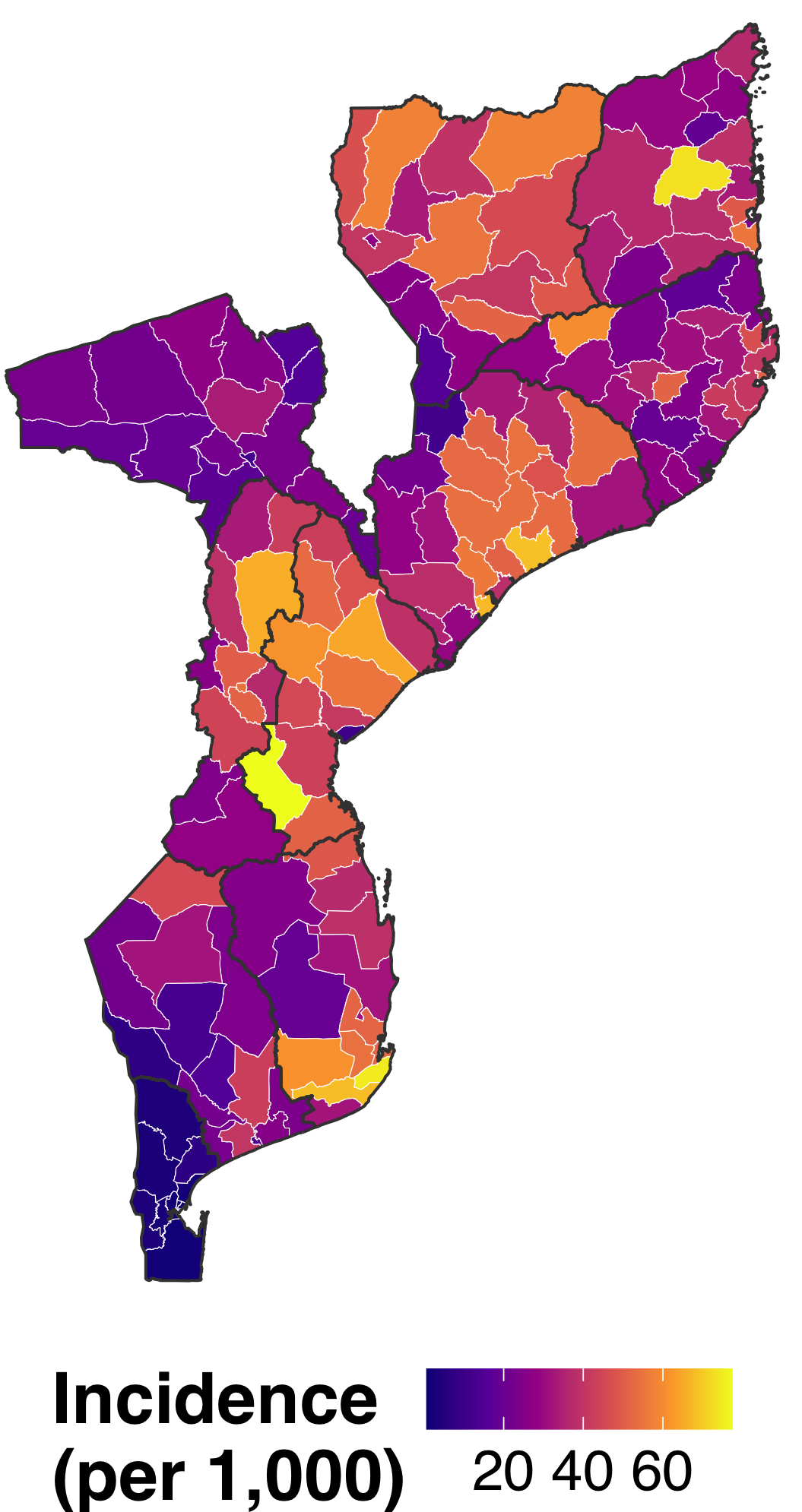}
    \caption{Spatial distribution of average monthly malaria incidence at district level in Mozambique from June 2017 to May 2024}
    \label{chp4.descript1}
\end{figure}

\begin{figure}
    \centering
    \includegraphics[width=1\linewidth]{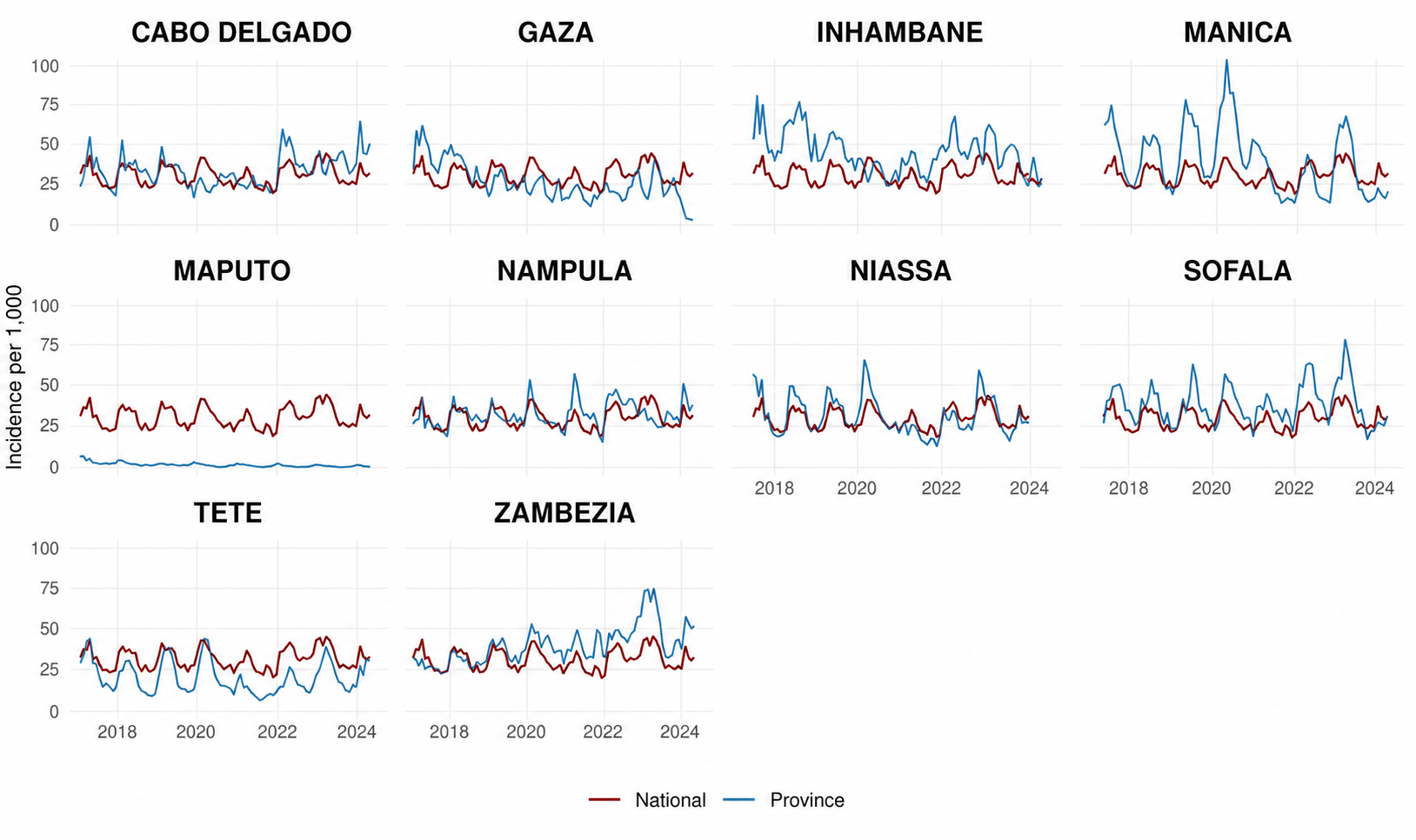}
    \caption{Temporal evolution of malaria incidence in Mozambique at both national and province level during the period June 2017 to May 2024}
    \label{fig:chap4_mapMalariatemp}
\end{figure}

The study was conducted in Mozambique, a malaria-endemic country in southeastern Africa, and covered the period from July 2017 to May 2024. The analysis combined daily environmental data with monthly counts of laboratory-confirmed malaria cases reported by health facilities to the National Malaria Control Program of Mozambique and provided under a Data Transfer Agreement. Malaria cases were aggregated at the district level, comprising 161 districts distributed across 10 provinces. District-level population estimates were obtained from the 2017 Population and Housing Census and its official projections \citep{ine2017census} and were used to calculate malaria incidence rates for descriptive analyses. Figure~\ref{chp4.descript1} presents the spatial distribution of average malaria incidence across the country, while Figure~\ref{fig:chap4_mapMalariatemp} shows its temporal evolution at both the national and provincial levels. \\\\

\subsection{Environmental data}
Daily climatic variables, including mean temperature, total precipitation, and mean relative humidity, were obtained from the NASA Prediction Of Worldwide Energy Resources (POWER) database developed by NASA’s Langley Research Center. In contrast to the monthly malaria surveillance data, these environmental variables were available at a daily temporal resolution. The POWER database combines satellite observations, atmospheric reanalysis products, data assimilation systems, and ground-based measurements to provide gridded climatic estimates. Data were retrieved using the \texttt{nasapower} package (version 4.2.5) in R \citep{nasapower}. These variables were extracted at the geographic centroid of each district.  \\  

Elevation data were obtained from the WorldClim global elevation dataset at a spatial resolution of 30 arc-seconds (approximately 1 km) using the R package \texttt{geodata} (version 0.6-2). District-level elevation was obtained by averaging raster values within each district boundary.  \\ 

The Normalized Difference Vegetation Index (NDVI), a remote sensing indicator of vegetation density and greenness, was obtained from the NASA AppEEARS platform \citep{AppEEARS2026} using MODIS vegetation index products (version 6.1) at a spatial resolution of 250 m \citep{MODISdata}. The MODIS product provides one NDVI observation approximately every 16 days, resulting in two observations per month. District-level NDVI values were calculated by averaging raster values within each district boundary, and monthly NDVI was defined as the maximum value observed during each month.  \\

Additional descriptive analyses are provided in Appendix~\ref{app:ch4_descrip}. These include the spatial distributions of average temperature, precipitation, relative humidity, elevation, and NDVI across Mozambique, as well as the temporal patterns of the climatic variables and NDVI in relation to national malaria incidence.

\subsection{Intervention data}
Information on malaria interventions, including Indoor Residual Spraying (IRS) coverage and Insecticide-treated Net (ITN) coverage, was obtained from the Malaria Atlas Project \citep{MAP2026}. These intervention indicators were available annually at the provincial level. Provincial estimates were assigned to all districts within the corresponding province and carried forward to all months within each year. Although these variables were available at a coarser spatial and temporal resolution than the malaria surveillance data, they were included to account for broad intervention-related differences in malaria transmission.

\subsection{Statistical methodology}

A hierarchical Bayesian modeling framework was developed to investigate the association between daily climatic exposures and monthly malaria incidence, explicitly accounting for the temporal misalignment between the observed malaria counts and the climatic covariates. The framework is therefore formulated at the daily level and linked to the observed monthly counts through an aggregation step. In addition, the model incorporates seasonal effects, malaria intervention variables, and latent temporal or spatio-temporal random effects to account for residual dependence and unexplained variability. \\

Let $Y_{it}$ denote the observed number of malaria cases in district $i=1,\ldots,N$ during month $t=1,\ldots,T$. It is assumed that the observed monthly counts arise from the aggregation of latent daily counts,
\begin{equation}
Y_{it}
=
\sum_{d\in\mathcal D_t}
Z_{itd},
\end{equation}
where $\mathcal D_t$ denotes the set of days belonging to month $t$, and $Z_{itd}$ is the latent number of malaria cases occurring on day $d = 1,\ldots,D_T$. This temporal disaggregation strategy is based on the proposal by \cite{fajgenblat2025temporal}, who implemented temporal disaggregation in the context of ecology. \\

Conditional on the latent daily intensity, the daily counts are assumed to be independent and follow a negative binomial distribution,

\begin{equation}
Z_{itd} | \mu_{itd}
\sim
\mathrm{NegBin}(\mu_{itd},\phi),
\end{equation}

where $\mu_{itd}$ denotes the expected number of malaria cases on day $d$ and $\phi$ is the overdispersion parameter. Under the following parameterization of a negative Binomial distribution,
\[
\mathbb E(Z_{itd})
=
\mu_{itd},
\qquad
\mathrm{Var}(Z_{itd})
=
\mu_{itd}
+
\frac{\mu_{itd}^{2}}{\phi},
\]
the aggregated monthly count also follows a negative binomial distribution,
\begin{equation}
Y_{it}
\sim
\mathrm{NegBin}(\lambda_{it},\phi),
\end{equation}
where the monthly mean is obtained by aggregating the latent daily expectations,
\begin{equation}
\label{ch4.eqaggre}
\lambda_{it} =
\sum_{d\in\mathcal D_t}
\mu_{itd}.
\end{equation}
The daily expected counts were modeled using a log link with population offset,
\begin{equation}
\label{ch4.dailymean}
\log(\mu_{itd})
=
\log(\mathrm{pop}_{it})
+
\eta_{itd},
\end{equation}
where $\mathrm{pop}_{it}$ denotes the district population and $\eta_{itd}$ is the daily linear predictor,
\begin{equation}
\eta_{itd}
=
\alpha_i
+
\eta_t^{\mathrm{seas}}
+
\eta_{itd}^{\mathrm{clim}}
+
\eta_{it}^{\mathrm{int}}
+
u_{it},
\label{eq:linear_predictor}
\end{equation}
where $\alpha_i$ denotes the district-specific baseline malaria risk, $\eta_t^{\mathrm{seas}}$ the seasonal component, $\eta_{itd}^{\mathrm{clim}}$ the climatic effect, $\eta_{it}^{\mathrm{int}}$ the intervention effect, and $u_{it}$ the latent temporal or spatio-temporal random effects. \\

Substituting the daily expected counts (Equation~\ref{ch4.dailymean}) into the monthly aggregation (Equation~\ref{ch4.eqaggre}) translates the monthly expected counts into,
\begin{align}
\lambda_{it}
&=
\sum_{d\in\mathcal D_t}
\exp\left(
\log(\mathrm{pop}_{it})
+
\eta_{itd}
\right).
\end{align}
An important feature of the proposed model is that only the climatic component varies within each month. Consequently,
\[
\eta_{itd}
=
\underbrace{
\alpha_i
+
\eta_t^{\mathrm{seas}}
+
\eta_{it}^{\mathrm{int}}
+
u_{it}
}_{\text{constant within month}}
+
\eta_{itd}^{\mathrm{clim}},
\]
which allows the monthly mean to be factorized as

\begin{align}
\lambda_{it}
&=
\mathrm{pop}_{it}
\exp\left(
\alpha_i
+
\eta_t^{\mathrm{seas}}
+
\eta_{it}^{\mathrm{int}}
+
u_{it}
\right)
\sum_{d\in\mathcal D_t}
\exp\left(
\eta_{itd}^{\mathrm{clim}}
\right).
\end{align}
Equivalently, the monthly linear predictor can be written as

\begin{equation}
\log(\lambda_{it})
=
\log(\mathrm{pop}_{it})
+
\alpha_i
+
\eta_t^{\mathrm{seas}}
+
\eta_{it}^{\mathrm{int}}
+
u_{it}
+
\log\left(
\sum_{d\in\mathcal D_t}
\exp\left(
\eta_{itd}^{\mathrm{clim}}
\right)
\right).
\label{eq:monthly_predictor}
\end{equation}

Equation~(\ref{eq:monthly_predictor}) shows that only the climatic contribution requires aggregation across days, while all remaining model components factor outside the summation. This formulation naturally accommodates daily climatic covariates and distributed lag nonlinear effects while preserving a likelihood defined for the observed monthly malaria counts.

\subsubsection{Seasonal effect}

To account for smooth annual seasonality in malaria incidence, the seasonal component was modeled using the first two harmonic terms with a period of 12 months,
\begin{equation}
\begin{aligned}
    \eta_t^{\mathrm{seas}} &= \beta_{\sin1} \sin\left(\frac{2\pi t}{12}\right) + \beta_{\cos1}
\cos\left(\frac{2\pi t}{12}\right) \\ 
&+  \beta_{\sin2} \sin\left(\frac{2\pi t}{6}\right) + \beta_{\cos2}
\cos\left(\frac{2\pi t}{6}\right).
\end{aligned}
\end{equation}
Weakly informative Gaussian priors were assigned to the harmonic coefficients,
\begin{equation*}
\beta_{\sin1},\beta_{\cos1}, \beta_{\sin2},\beta_{\cos2}
\sim
\mathcal{N}(0,0.5^2).
\end{equation*}

\subsubsection{Climatic effect}\label{chp4.climatic.effect}

The climatic component comprised four time-varying environmental variables: daily mean temperature, daily total precipitation, daily mean relative humidity, and monthly Normalized Difference Vegetation Index (NDVI), together with district-level average elevation. Since the time-varying climatic variables exhibit both spatial and temporal variation, each covariate was decomposed into a \textit{long-term spatial mean} (between-district component) and a \textit{temporal anomaly} (within-district component), following the between-within decomposition proposed by \citet{betw_with_decomp1},

\begin{equation}
X_{itd}^{(c)}
=
\overline X_i^{(c)}
+
\left(
X_{itd}^{(c)}
-
\overline X_i^{(c)}
\right),
\end{equation}
where $c$ indexes the climatic variable. The \textit{long-term spatial mean}, $\overline X_i^{(c)}$, represents the average exposure in district $i$ over the study period and therefore captures persistent differences in climatic conditions between districts. The \textit{temporal anomaly}, $X{itd}^{(c)}-\overline X_i^{(c)}$, measures the deviation from the district-specific long-term mean at a given time and therefore captures short-term variation in exposure within each district. This decomposition separates persistent spatial differences from temporal departures from typical local conditions, thereby reducing potential ecological and spatial confounding. \\

The \textit{long-term spatial means} were included linearly in the predictor, whereas the \textit{temporal anomalies} were modeled using a Distributed Lag Nonlinear Model (DLNM) \citep{gasparrini2010dlnm}, allowing both nonlinear exposure-response relationships and delayed lag effects. For climatic variable $c$, the within-district contribution is given by

\begin{equation}
f_c\!\left(
X_{itd}^{(c)}
-
\overline X_i^{(c)}
\right)
=
\sum_{j=1}^{J_c}
\sum_{m=1}^{M_c}
\left[
\sum_{\ell=0}^{L_c}
B_j^{(c)}
\!\left(
X_{i,t-\ell,d}^{(c)}
-
\overline X_i^{(c)}
\right)
H_m^{(c)}(\ell)
\right]
\beta_{jm}^{(c)},
\end{equation}

where $B_j^{(c)}(\cdot)$ and $H_m^{(c)}(\cdot)$ denote the exposure-response and lag-response basis functions, respectively, $\beta_{jm}^{(c)}$ are the cross-basis coefficients, $L_c$ is the maximum lag, and $J_c$ and $M_c$ denote the numbers of basis functions in the exposure and lag dimensions.\\

Natural cubic splines were used for both the exposure-response and lag-response functions, with knots placed at equally spaced quantiles of the observed exposure and lag distributions. For mean daily temperature, total precipitation, and mean daily relative humidity, the maximum lag was set to $L_c=59$ days to account for delayed effects associated with mosquito development, parasite incubation, and malaria transmission. For NDVI, the maximum lag was set to six months to reflect slower ecological processes influencing malaria transmission. The exposure and lag dimensions for temperature, precipitation, and humidity were represented using $J_c=M_c=3$ basis functions, whereas for NDVI the exposure dimension used $J_c=3$ basis functions and the lag dimension used $M_c=2$ basis functions.\\

Furthermore, elevation was added linearly into the model. The climatic component of the linear predictor is therefore given by

\begin{equation}
\begin{aligned}
\eta_{itd}^{\mathrm{clim}}
=
&
\beta_{\mathrm{temp}}
\overline X_i^{\mathrm{temp}}
+
\beta_{\mathrm{prec}}
\overline X_i^{\mathrm{prec}}
+
\beta_{\mathrm{hum}}
\overline X_i^{\mathrm{hum}}
+
\beta_{\mathrm{NDVI}}
\overline X_i^{\mathrm{NDVI}}
\\
&
+
f_{\mathrm{temp}}
\left(
X_{itd}^{\mathrm{temp}}
-
\overline X_i^{\mathrm{temp}}
\right)
+
f_{\mathrm{prec}}
\left(
X_{itd}^{\mathrm{prec}}
-
\overline X_i^{\mathrm{prec}}
\right)
\\
&
+
f_{\mathrm{hum}}
\left(
X_{itd}^{\mathrm{hum}}
-
\overline X_i^{\mathrm{hum}}
\right)
+
f_{\mathrm{NDVI}}
\left(
X_{it}^{\mathrm{NDVI}}
-
\overline X_i^{\mathrm{NDVI}}
\right)
\\
&
+\beta_{\mathrm{elev}}
X_i^{\mathrm{elev}}.
\end{aligned}
\end{equation}

The cross-basis coefficients were assigned hierarchical non-centered priors,
\begin{equation*}
\beta_{jm}^{(c)}
=
\tau_c
\theta_{jm}^{(c)},
\qquad
\theta_{jm}^{(c)}
\sim
\mathcal N(0,1),
\end{equation*}
where $\tau_c$ controls the overall magnitude of the spline coefficients and was assigned the prior
\[
\tau_c
\sim
\mathrm{half\text{-}normal}(0,0.2).
\]
Weakly informative Gaussian priors,
\[
\beta_{\mathrm{temp}},
\beta_{\mathrm{prec}},
\beta_{\mathrm{hum}},
\beta_{\mathrm{NDVI}},
\beta_{\mathrm{elev}}
\sim
\mathcal N(0,0.5^2),
\]
were assigned to the linear effects associated with the between-district climatic components and elevation.

\subsubsection{Intervention effect}

The intervention component accounts for the effects of malaria control strategies on disease transmission. Specifically, Indoor Residual Spraying (IRS) coverage and Insecticide-treated Net (ITN) coverage were included as linear effects,
\begin{equation}
\eta_{it}^{\mathrm{int}}
=
\beta_{\mathrm{ITN}}X_{it}^{\mathrm{ITN}}
+
\beta_{\mathrm{IRS}}X_{it}^{\mathrm{IRS}},
\end{equation}
where $X_{it}^{\mathrm{ITN}}$ and $X_{it}^{\mathrm{IRS}}$ denote district-level ITN and IRS coverage, respectively.\\

Weakly informative Gaussian priors were assigned to the intervention coefficients,

\[
\beta_{\mathrm{ITN}},
\beta_{\mathrm{IRS}}
\sim
\mathcal{N}(0,0.5^2).
\]

\subsubsection{Random effects}

The random-effects component accounts for residual temporal dependence not explained by the observed covariates. Following the approach proposed by \citet{MALE}, each district was assigned a temporal Gaussian process,
\begin{equation*}
\mathbf u_i
=
(u_{i1},\ldots,u_{iT})^\top
\sim
\mathcal N(\mathbf0,K_i),
\end{equation*}

where $K_i$ is a Matérn covariance matrix with smoothness parameter fixed at $\nu=1.5$. The covariance between two time points $t$ and $t'$ is
\begin{equation*}
K_i(t,t')
=
\sigma_i^2
\left(
1+\frac{\sqrt3|t-t'|}{\rho}
\right)
\exp\left(
-\frac{\sqrt3|t-t'|}{\rho}
\right),
\end{equation*}

where $\sigma_i^2$ controls the marginal temporal variability for district $i$, while $\rho$ determines the rate at which temporal correlation decays. Prior to model fitting, the temporal domain was rescaled to the interval $[-1,1]$.\\

Rather than assuming a common marginal variance across districts, district-specific variance parameters were assigned a spatially structured conditional autoregressive prior,
\begin{equation*}
\sigma_i^2
\mid
\sigma_{-i}^2
\sim
\mathcal N
\left(
\frac1{n_i}
\sum_{j\sim i}
\sigma_j^2,
\frac{\tau_\sigma^2}{n_i}
\right),
\end{equation*}

where neighboring districts were defined using a Queen contiguity adjacency matrix. This specification encourages neighboring districts to exhibit similar levels of temporal variability while allowing substantial regional heterogeneity in the magnitude of temporal fluctuations.\\

The temporal correlation range was assigned the prior
\[
\rho
\sim
\mathrm{Inverse\text{-}Gamma}(5,5).
\]
As sensitivity analyses, four alternative random-effects structures were also considered: (i) no random effects, (ii) a first-order random walk, (iii) a temporal Gaussian process with a common marginal variance across districts, and (iv) the Knorr--Held \citep{knorr2000bayesian} spatio-temporal interaction model. Details of these alternative specifications are provided in the ~\ref{app:ch4_param_re}.

\subsubsection{Model implementation, selection and performance evaluation}

All models were implemented in Stan using the R package \texttt{cmdstanr} (version 0.8.1). Posterior inference was performed using Hamiltonian Monte Carlo (HMC) with 500 iterations burn-in and 500 of sampling. Convergence and sampling quality were assessed using trace plots, autocorrelation plots, the potential scale reduction statistic ($\hat{R}$), effective sample sizes (ESS), and the inspection of divergent transitions. \\

The monthly likelihood requires the aggregation of latent daily expected counts, to reduce the computational burden associated with repeatedly evaluating this quantity during posterior sampling, we approximated the aggregation term using a second-order Taylor expansion. Details of the approximation are provided in the ~\ref{app:ch4_compu_impl}.\\

Model fit was assessed using the Watanabe-Akaike Information Criterion (WAIC) \citep{watanabe2010asymptotic}. As a benchmark, an analogous DLNM without temporal disaggregation was additionally fitted, in which daily climatic variables were first aggregated to the monthly scale before model fitting. For this benchmark model, the maximum lag was set to four months for mean temperature, total precipitation, and relative humidity, and six months for NDVI. The exposure-response and lag-response functions were represented using natural cubic splines with $J_c=3$ basis functions for the exposure dimension and $M_c=2$ basis functions for the lag dimension.\\

Predictive performance was evaluated using a temporal hold-out strategy. The data were divided into a training set (July 2017-October 2023) and a test set (November 2023-April 2024). Models were estimated using the training data, and out-of-sample predictions were evaluated on the test data using the Root Mean Squared Error (RMSE) and the Empirical Coverage Probability (ECP). Whereas WAIC was used to compare in-sample model fit while accounting for model complexity, RMSE and ECP were used to assess predictive accuracy and uncertainty calibration on unseen observations.

\section{Results}\label{ch4_results}
Convergence diagnostics indicated satisfactory convergence for both the monthly model and the model incorporating temporal disaggregation. In general, $\widehat{R}$ values were close to 1 across model parameters, while visual inspection of the trace plots indicated adequate mixing and no apparent convergence issues. Table~\ref{tab:chap4_performance} summarizes the predictive performance of the proposed temporally disaggregated model and an otherwise identical model in which the climatic variables were first aggregated to monthly values. The proposed framework achieved a substantially lower WAIC (197,379) than the model without temporal disaggregation (198,422), indicating a better balance between model fit and complexity. Consistent with this result, the temporally disaggregated model also obtained lower RMSE values and improved empirical coverage probabilities (ECP) in both the training and test datasets, demonstrating that explicitly accounting for the daily variability of climatic exposures improves both predictive accuracy and uncertainty quantification. 

\begin{table}[ht]
\centering
\begin{tabular}{lcccc}
\hline
\textbf{Model} &
\multicolumn{2}{c}{\textbf{Training}} &
\multicolumn{2}{c}{\textbf{Test}} \\
\cline{2-5}
& \textbf{RMSE} & \textbf{ECP} & \textbf{RMSE} & \textbf{ECP} \\
\hline
Without temporal disaggregation & 1039 & 0.99 & 3602 & 0.86 \\
With temporal disaggregation    & 988 & 0.99 & 3284 & 0.88 \\
\hline
\end{tabular}
\caption{Comparison of the predictive performance of the proposed temporally disaggregated model and an otherwise identical model using monthly aggregated climatic variables. RMSE denotes the root mean squared error and ECP the empirical coverage probability, both evaluated on the training and test datasets.}
\label{tab:chap4_performance}
\end{table}

\ref{app:ch4_sensitivity} compares the proposed model with alternative random-effects specifications. Overall, the specification with spatially structured variance parameters for the Gaussian processes \citep{MALE} consistently provided the best model fit, supporting its suitability for capturing the complex spatio-temporal variability of malaria risk. \\


The ECP results further demonstrated that the improved predictive accuracy was achieved without substantially compromising uncertainty quantification. For the training data, ECP values were nearly identical for both models across all provinces, remaining consistently above the nominal 95\% level and therefore indicating slightly conservative predictive intervals. In the test data, the temporally disaggregated model generally maintained or improved empirical coverage compared with the model using monthly aggregated climatic variables. \\

\begin{figure}[h]
    \centering
    \includegraphics[width=1\linewidth]{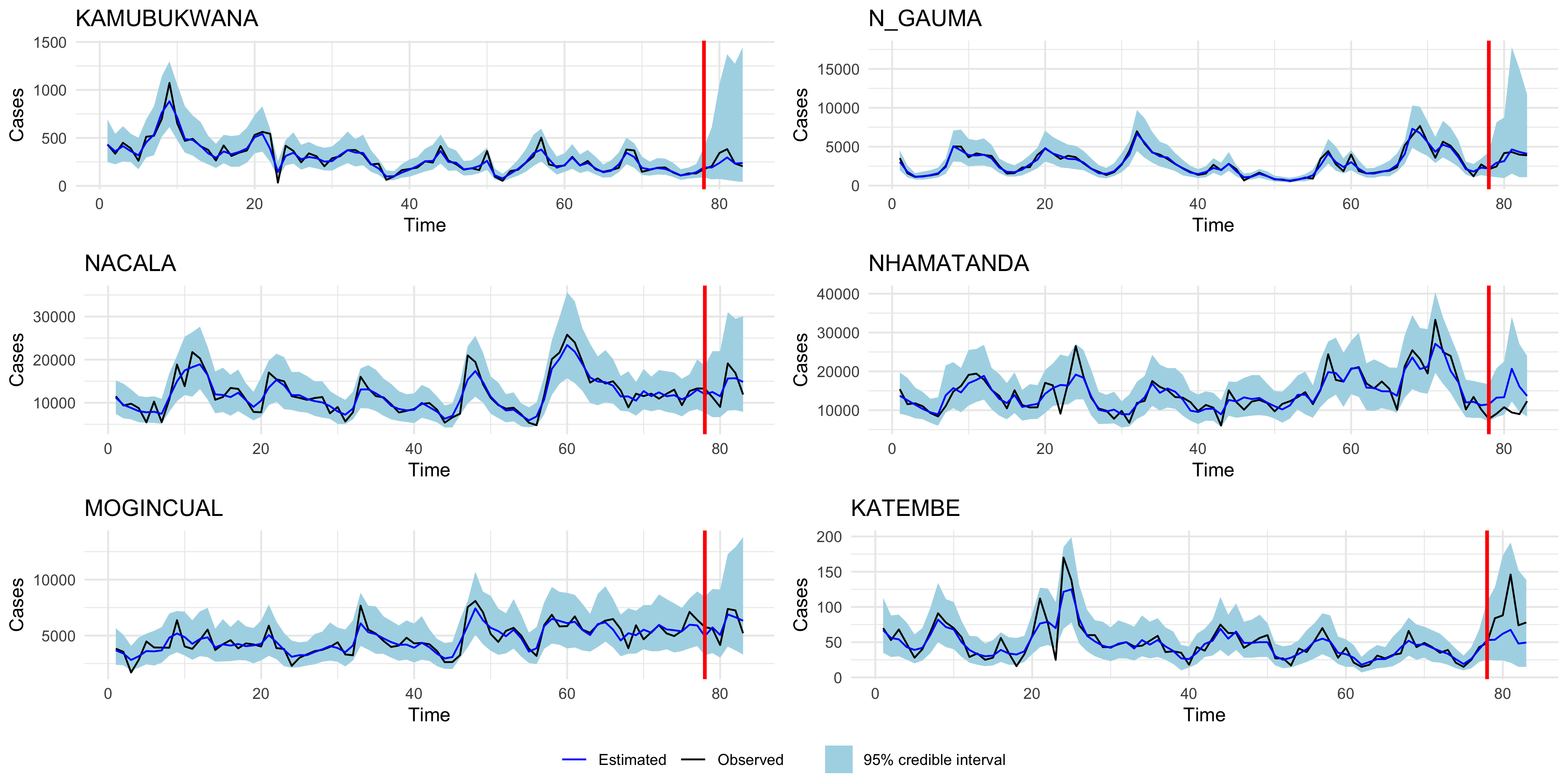}
    \caption{Observed and predicted monthly malaria cases for six randomly selected districts. The vertical red line indicates the division between the training and test sets.}
    \label{fig:chap5_predictions}
\end{figure}

Figure \ref{fig:chap5_predictions} illustrates the predictive performance of the proposed model for six randomly selected districts. Overall, the model successfully captures the main temporal patterns in malaria incidence, including seasonal fluctuations and several major epidemic peaks. For most districts, the observed counts remain within the 95\% posterior credible intervals throughout both the training and test periods, consistent with the relatively high empirical coverage probabilities (ECP) reported in Table \ref{tab:chap4_performance}. As expected, predictive uncertainty increases after the train-test split, as reflected by the wider credible intervals in the forecasting period. While the model reproduces the general temporal dynamics well, some abrupt peaks and local fluctuations are either underestimated or overestimated, particularly in districts with highly variable incidence patterns. \\

Posterior summaries of the proposed model are presented in Table \ref{tab:posterior_summary}. The \textit{long-term spatial mean} effects of mean temperature, NDVI, and elevation were all positively associated with malaria incidence, with 95\% credible intervals excluding zero. Among these covariates, mean temperature exhibited the strongest association ($\beta_{\text{temp}}=1.16$, 95\% CI: 0.83, 1.45), corresponding to a relative risk (RR) of approximately 3.19. NDVI and elevation were also positively associated with malaria incidence, with estimated RRs of approximately 2.32 and 1.60, respectively. Both intervention variables exhibited negative associations with malaria incidence. While the estimated reductions in malaria risk were modest, the posterior distributions provided evidence of a protective effect, with the 95\% credible intervals excluding zero for both intervention variables. Covariates whose 95\% credible intervals included zero were retained in the model to adjust for potential confounding and to account for established determinants of malaria transmission, rather than being removed solely on the basis of statistical significance. 

\begin{table}[ht]
\centering
\begin{tabular}{lrrrrrr}
\hline
Variable  & Median & SD & 2.5\% & 97.5\% & $\hat{R}$ & ESS Bulk \\
\hline
$\beta_{\text{temp}}$ & 1.170 & 0.165 & 0.832 & 1.450 & 1.01 & 129 \\
$\beta_{\text{prec}}$ & 0.021 & 0.327 & -0.539 & 0.645 & 1.03 & 122 \\
$\beta_{\text{humi}}$ & 0.182 & 0.169 & -0.117 & 0.493 & 1.03 & 111 \\
$\beta_{\text{ndvi}}$ & 0.848 & 0.116  & 0.598 & 1.070 & 1.00 & 103 \\
$\beta_{\text{elev}}$ & 0.473 & 0.080  & 0.321 & 0.616 & 1.03 & 126 \\
$\beta_{\text{itn}}$ & -0.00395 & 0.00063 & -0.00500 & -0.00264 & 1.01 & 442 \\
$\beta_{\text{irs}}$ & -0.00462 & 0.00126  & -0.00687 & -0.00233 & 0.999 & 215 \\
$\phi$ & 33.400 & 0.907 & 31.800 & 35.100 & 1.01 & 19.1 \\
\hline
\end{tabular}
\caption{Posterior summaries of model parameters}
\label{tab:posterior_summary}
\end{table}

\begin{figure}[ht]
\centering
\includegraphics[width=\linewidth]{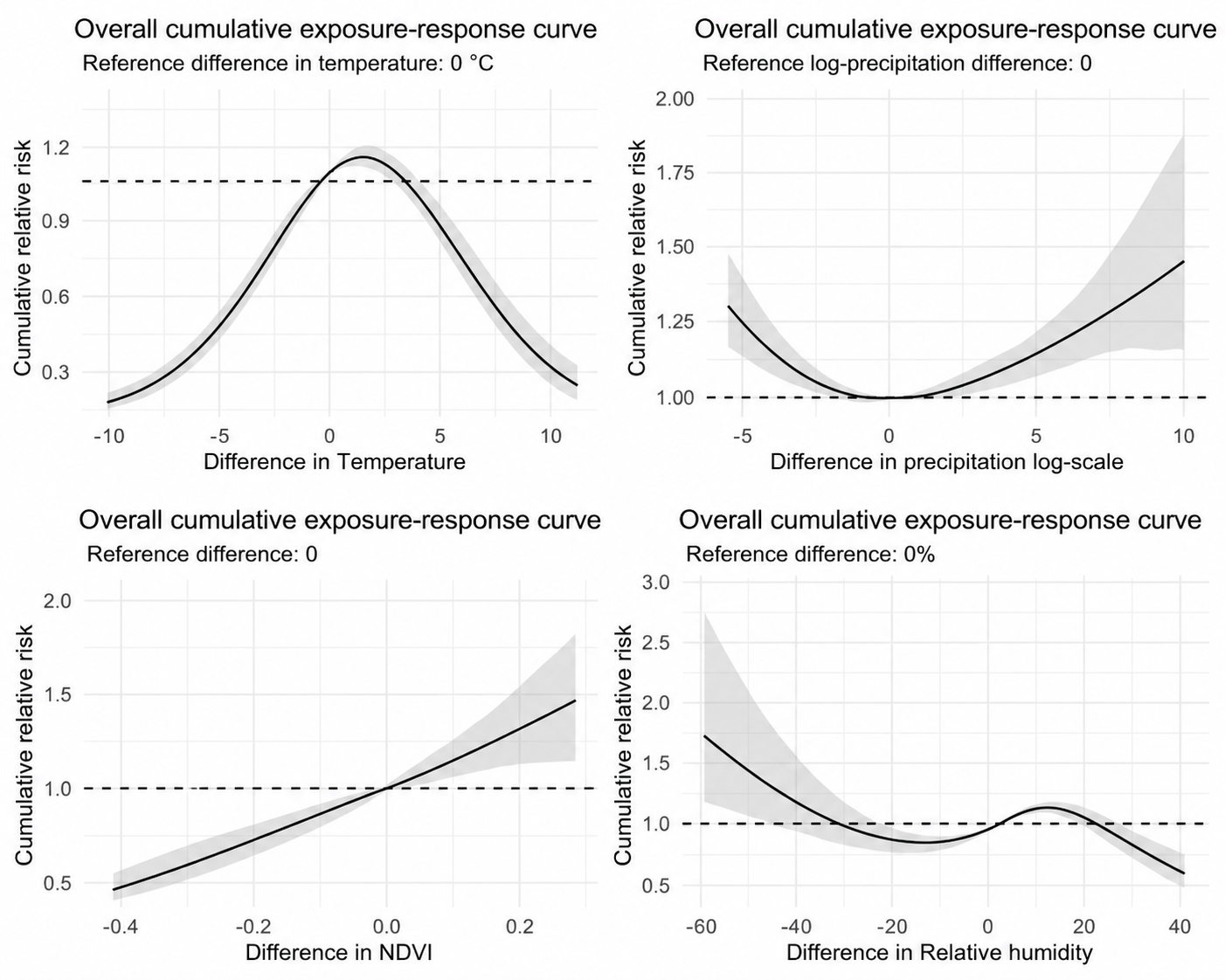}
\caption{Estimated cumulative exposure-response curves for the four environmental variables.}
\label{fig:chap5_er_climate_effects}
\end{figure}

\begin{figure}[!b]
\centering
\includegraphics[width=\linewidth]{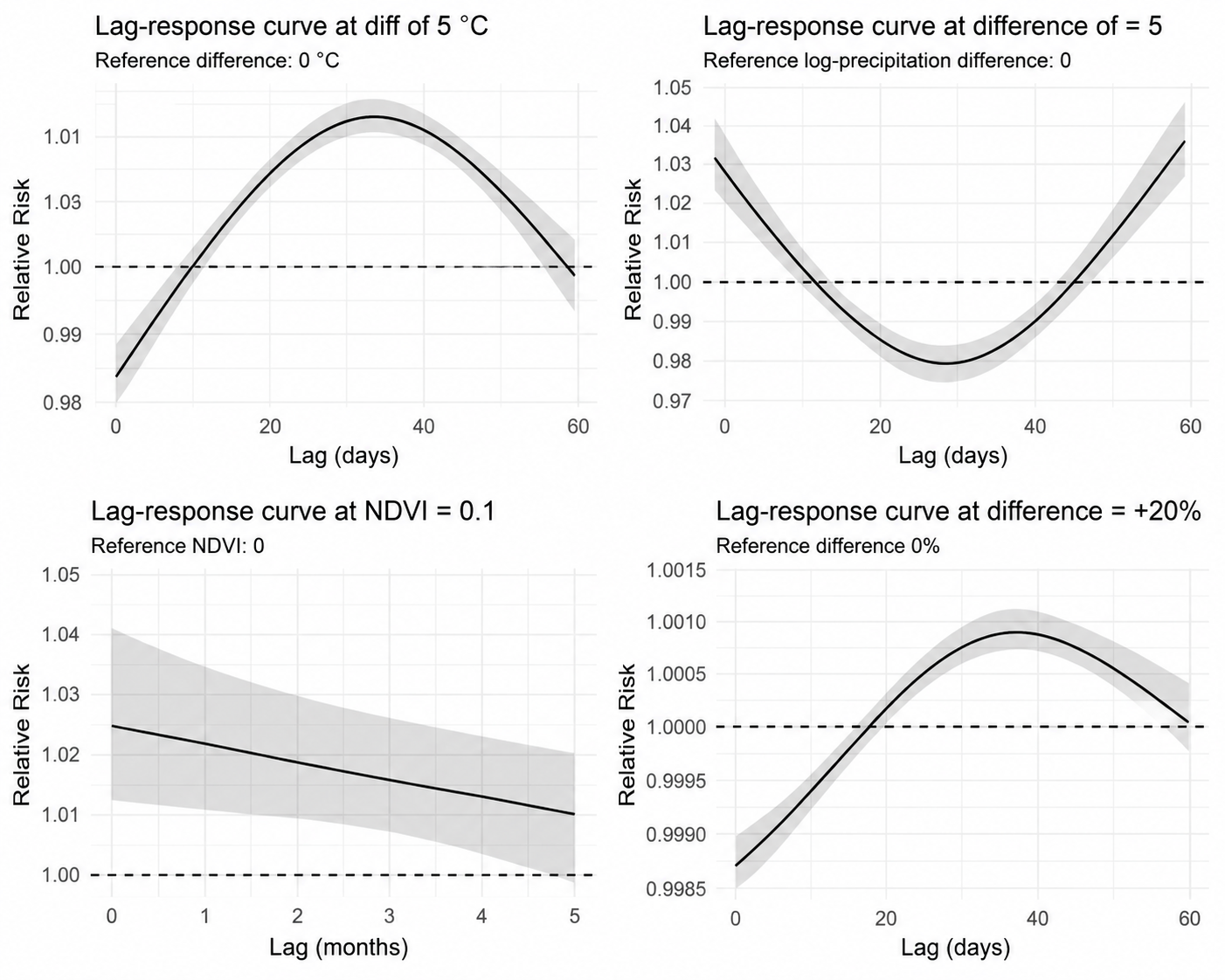}
\caption{Estimated lag-response curves for the four environmental variables.}
\label{fig:chap5_lr_all_climate_effects}
\end{figure}

Figures~\ref{fig:chap5_er_climate_effects} and \ref{fig:chap5_lr_all_climate_effects} summarize the estimated nonlinear exposure-response and lag-response relationships for the climatic \textit{temporal anomalies}. As described in Section~\ref{chp4.climatic.effect}, the climatic variables were decomposed into \textit{long-term spatial means} and \textit{temporal anomalies} components. The distributed lag nonlinear approach was applied only to the later, therefore, the estimated effects should be interpreted as the impact of temporal deviations from each district’s long-term average climatic conditions, rather than absolute exposure values. For all variables, the reference exposure corresponds to no deviation from the district-specific average (i.e., an anomaly of zero), such that positive and negative values represent conditions above and below the local climatic baseline, respectively.\\ 

Figure~\ref{fig:chap5_er_climate_effects} presents the cumulative exposure-response relationships for the climatic \textit{temporal anomalies}. Overall, the estimated associations were nonlinear, although their shapes differed across environmental variables. Panel (a) shows that both negative and positive temperature anomalies were associated with lower cumulative relative risk than the reference condition, suggesting that malaria transmission is favored within a relatively narrow range around the district-specific average temperature. Panel~(b) exhibits a nonlinear U-shaped association between log-transformed precipitation anomalies and malaria risk, with the lowest cumulative relative risk occurring near the reference condition and higher risks observed for both negative and positive precipitation anomalies. Uncertainty increased at the extremes, particularly for large positive anomalies, likely reflecting the limited number of observations at these exposure levels. 
Panel~(c) indicates an approximately monotonic positive association between NDVI anomalies and malaria incidence, with cumulative relative risk increasing as vegetation conditions became greener than the district-specific average. Finally, panel~(d) reveals a nonlinear association between relative humidity anomalies and malaria risk. Moderate positive anomalies were associated with a slight increase in relative risk, whereas both large negative and large positive anomalies were associated with lower risk. \ref{app:ch4_betwith} presents the distributions of the within-district anomalies, together with the district-specific average values (i.e., the between-district component), allowing the exposure-response curves to be interpreted in the context of both the observed range of anomalies and the local climatic conditions against which they are defined.\\ 

Figure \ref{fig:chap5_lr_all_climate_effects} presents the cumulative lag-response curves describing the association between the within-district climatic component and malaria incidence. The lag-response curves were evaluated at fixed positive differences maintained throughout the lag period: a $5^\circ\text{C}$ increase for temperature, a positive difference of 5 units in log-transformed precipitation, a $20\%$ increase in relative humidity, and an NDVI anomaly of 0.1. Overall, the within-district components of temperature, precipitation, and relative humidity exhibited nonlinear lag-response, whereas the association with NDVI appeared approximately linear over the range of observed values. Panel (a) suggests that for a $5^\circ\text{C}$ temperature increase the effect is delayed, peaking approximately 25-35 days after exposure before gradually declining toward the null effect. Panel (b) indicates that for a positive precipitation difference of 5 units on the log scale there is a significant delayed effect, with malaria risk increasing gradually and peaking approximately 35-45 days after exposure before stabilizing. Panel (c) shows that for a positive $20\%$ relative humidity difference there is a delayed but relatively small effect, with malaria risk peaking approximately 30-40 days after exposure before gradually returning toward the null effect. Panel (d) presents the lag-response curve for an NDVI anomaly of $0.1$ which suggests that the effect is relatively persistent over time, with elevated malaria risk observed immediately after exposure and gradually decreasing over subsequent months.

\section{Discussion}\label{ch4_discussion}

This study developed a Bayesian spatio-temporal framework for disease mapping that enables high-resolution environmental data to be incorporated into the analysis of  lower-resolution epidemiological data. The proposed methodology integrates temporal disaggregation within a comprehensive disease mapping framework that simultaneously accounts for delayed and nonlinear environmental effects, as well as residual spatial and temporal heterogeneity. By adopting a Bayesian hierarchical formulation, the framework allows the joint estimation of all model component while propagating uncertainty throughout the inferential process. The methodology was illustrated using malaria surveillance data from Mozambique, where daily environmental information was linked to monthly malaria cases counts between June 2017 to May 2024. Compared with models based on monthly aggregated environmental exposures, the proposed framework consistently achieved better predictive performance, demonstrating the practical value of explicitly modeling the underlying disease process at a finer temporal resolution.\\

The temporal disaggregation strategy used in this study provides an alternative approach for addressing temporal misalignment between environmental exposures and health outcomes, complementing the methodologies recently proposed by  \cite{economou2026modelling} and \cite{shukla2026mixed}. Like the approach of \cite{economou2026modelling}, the proposed framework exploits the infinite divisibility property of probability distributions. \\

However, whereas their method is based on the Poisson distribution, our framework uses the negative binomial distribution under a parameterization for which this property also holds. Unlike the formulation of \cite{economou2026modelling}, the proposed framework retains the canonical link function while reconstructing a latent daily disease process that is subsequently aggregated to the observed monthly case counts. By modeling the disease process at the temporal scale at which transmission occurs, the framework preserves the short-term environmental variability that may be obscured by conventional temporal aggregation. \\

Compared with the mixed-frequency DLNM proposed by \cite{shukla2026mixed}, the proposal establishes the relationship between environmental exposures and disease outcomes at a common temporal resolution, rather than estimating lagged associations between variables observed at different temporal scales. Nevertheless, a formal comparison of these three approaches through simulation studies would provide valuable insights into their relative advantages, computational efficiency, and inferential performance of each methodology under different data-generating mechanisms. \\

An additional strength of the proposed framework is the decomposition of time-varying climatic variables into between- and within-district components \citep{betw_with_decomp1}. By estimating climatic effects from temporal deviations around each district’s long-term average, the approach reduces potential spatial confounding from time-invariant differences between districts while providing a locally referenced interpretation of the nonlinear exposure-response relationships. This is particularly relevant in Mozambique, where identical absolute climatic conditions may have different ecological implications across regions with distinct baseline climates. Although climatic anomalies have been considered in previous malaria studies \citep{chirombo2020childhood, lake2026climate}, their integration with within-between decomposition, nonlinear distributed-lag modeling, and temporal disaggregation remains limited. The maps in \ref{app:ch4_betwith} further illustrate the spatial variation in baseline climatic conditions and the corresponding within-district anomalies. \\

Beyond its predictive performance, the proposed framework identified epidemiologically plausible nonlinear and delayed associations between climatic variability and malaria incidence. Malaria risk was highest at temperatures close to district-specific averages, while positive NDVI anomalies were associated with increasing risk, consistent with established effects of temperature and vegetation on malaria transmission \citep{yamba2023climate, ekpa2023spatio}. Both unusually low and high precipitation were associated with increased risk, potentially reflecting the formation of suitable breeding habitats as water bodies recede or following increased rainfall \citep{okiring2021associations, armando2025spatio}. Temperature and precipitation effects accumulated gradually over several weeks, consistent with the time required for environmental changes to influence vector breeding, development, parasite dynamics, and subsequent human infection. Nevertheless, these results represent adjusted epidemiological associations rather than causal effects, as residual confounding, measurement error, collinearity among environmental variables, and changes in surveillance or healthcare access cannot be excluded. \\ 

The proposed framework has important implications for malaria surveillance because it enables routinely collected health data to be integrated with high-resolution environmental information without requiring changes to existing surveillance systems. In many malaria-endemic countries, monthly district-level case counts remain the primary source of epidemiological information, whereas meteorological observations and satellite-derived environmental data are increasingly available at daily resolution. By explicitly addressing this temporal mismatch, the proposed methodology allows surveillance systems to exploit substantially more environmental information than approaches based on monthly summaries, thereby improving disease-risk estimation while remaining fully compatible with existing reporting practices. Although the framework is illustrated using malaria surveillance in Mozambique, its applicability extends to other climate-sensitive infectious diseases, such as dengue, Zika, and chikungunya, as well as to other epidemiological outcomes for which environmental or other explanatory variables are available at finer temporal resolutions than the health data. More broadly, the proposed approach aligns with the increasing emphasis on strengthening surveillance systems through the integration of diverse data sources and advanced analytical methods, as highlighted in the WHO report \textit{Future surveillance for epidemic and pandemic diseases: a 2023 perspective} \citep{world2023future}.  \\

Additionally, the increasing frequency of climate anomalies highlights the relevance of flexible approaches for characterizing environmental effects on malaria transmission. By allowing nonlinear and delayed effects, the proposed framework can help investigate how unusual climatic conditions, including those associated with El Niño, may influence malaria risk over subsequent weeks. This is increasingly relevant under climate change, although predictions under future conditions should be interpreted cautiously when they extend beyond the climatic range observed during the study period. \\

Despite its advantages, the proposed framework has limitations related to data quality and spatial resolution. Routine malaria surveillance may be affected by incomplete reporting, diagnostic practices, healthcare access, and reporting systems, with documented discrepancies in Mozambique \citep{colborn2020quality}. Although not explicitly considered here, reporting uncertainty could be incorporated by jointly modeling disease incidence and the reporting process within the hierarchical framework \citep{stoner2019hierarchical}. Remotely sensed climatic data are also subject to measurement uncertainty and may not capture the fine-scale conditions relevant to mosquito–environment interactions \citep{mcmahon2022comparing}. While spatio-temporal random effects account for residual heterogeneity, they cannot distinguish genuine variation in malaria transmission from reporting, measurement, or other unmeasured processes.\\

A further limitation is the potential multicollinearity among the environmental variables. Climatic variables are inherently correlated, particularly when considering long-term climatic averages, although the correlations among the within-district anomaly components were generally weaker and varied across districts (~\ref{app:ch4_coll}). Sensitivity analyses indicated that the estimated exposure-response relationships were largely robust to the joint inclusion of multiple climatic variables, with the main difference observed for the lag-response function of precipitation. Nevertheless, the estimated effects should be interpreted as conditional on the other variables included in the model rather than as the total effects of individual environmental drivers within the broader climate–malaria system.\\

Despite the fact that the proposed framework generally achieved good predictive calibration, empirical coverage probabilities were occasionally below the nominal 95\% level, indicating a slight underestimation of predictive uncertainty in some districts or time periods. This under-coverage may reflect residual sources of uncertainty that were not fully captured by the model, for instance, some local transmission dynamics, variability in routine surveillance data, measurement error in the environmental covariates, or limitations associated with the temporal disaggregation approach. Consequently, while the predictive intervals were generally well calibrated, there remains scope for further improving uncertainty quantification. \\

The decomposition into within- and between-district components also assumes that the effects of climatic anomalies are common across districts. In practice, however, the impact of temperature and precipitation anomalies is likely to vary according to local climatic conditions, vector ecology, land cover, intervention coverage, and baseline transmission intensity. Consequently, the estimated exposure-response functions should be interpreted as average associations across Mozambique rather than district-specific relationships, potentially masking meaningful spatial heterogeneity. Extending the proposed framework to allow spatially varying exposure-response functions represents an important direction for future research, for instance, exploring the spatially varying distributed lag non-linear models proposed by \cite{rutten2026spatially}. Likewise, the intervention variables were limited by their spatial and temporal resolution. As a result, the estimated intervention effects may be influenced by measurement error, intervention targeting, or limited temporal variability, and should therefore be interpreted with caution.\\

This study is based on an observational ecological design, and the estimated exposure-response and lag-response relationships should therefore be interpreted as adjusted epidemiological associations rather than causal effects. Environmental epidemiology is particularly susceptible to multiple sources of confounding because climatic variables, malaria transmission, and control activities vary simultaneously across space and time. To mitigate these sources of bias, the proposed framework incorporates several methodological features. Seasonal confounding was addressed through explicit modeling of seasonal patterns, while district-specific intercepts, elevation, and the decomposition of climatic variables into between- and within-district components accounted for persistent spatial differences. Temporal random effects captured long-term trends that might otherwise be attributed to climatic covariates, and available intervention indicators were included to adjust for malaria control activities. Nevertheless, residual confounding cannot be excluded. Environmental variables may still act as proxies for broader ecological processes, such as breeding habitat availability or land-use characteristics, while unmeasured changes in intervention implementation, healthcare access, socioeconomic conditions, or reporting practices may also influence the estimated associations. Overall, by jointly accounting for seasonal patterns, spatial heterogeneity, temporal trends, intervention measures, and delayed nonlinear environmental effects within a single hierarchical framework, the proposed methodology addresses several important sources of confounding commonly encountered in environmental epidemiology, although it cannot eliminate them entirely. \\

Several methodological extensions could further increase the flexibility and applicability of the proposed disease mapping framework. Although intervention effects were included in the current model, they were represented using relatively simple linear terms. Future work could model them as delayed, nonlinear, and time-varying processes, allowing the impact of malaria control measures to evolve over time, as proposed by \cite{giorgi2025decay}. Another promising direction is the incorporation of multivariate distributed lag models to represent compound environmental effects arising from the joint action of temperature, precipitation, humidity, and vegetation \citep{fletcher2025compound}. Together, these developments would provide a more realistic representation of the complex mechanisms underlying malaria transmission, while preserving the advantages of the proposed Bayesian hierarchical framework. More broadly, these extensions would also facilitate the application of the framework to other infectious diseases and epidemiological outcomes influenced by environmental and climatic conditions. 

\section{Conclusion}\label{ch4_conclu}

This paper proposed a Bayesian spatio-temporal modeling framework that integrates temporal disaggregation, distributed lag nonlinear models, and spatial random effects to jointly analyze monthly disease surveillance data and daily environmental exposures. By modeling the disease process at the daily scale while preserving monthly observations, the framework enables the use of high-resolution environmental information without requiring temporal aggregation. Applied to malaria surveillance in Mozambique, this approach improved predictive performance relative to conventional monthly models while recovering biologically plausible exposure-response and lag-response relationships. More broadly, the methodology provides a flexible framework for studying delayed and nonlinear environmental effects in settings where health and exposure data are observed at different temporal resolutions. Its applicability extends beyond malaria to other environmentally driven diseases and surveillance systems with temporally misaligned data.

\bibliographystyle{elsarticle-harv} 
\bibliography{elsarticle-3/bibliography.bib}

\newpage

\appendix

\section{Exploratory descriptive plots of covariates information} \label{app:ch4_descrip}

This appendix provides additional descriptive analyses of the environmental variables considered in the study, including temperature, precipitation, relative humidity, elevation, and the Normalized Difference Vegetation Index (NDVI). Both spatial and temporal patterns are presented to characterize environmental variability across Mozambique during the study period.\\

\begin{figure}[!b]
\centering
\includegraphics[width=0.8\linewidth]{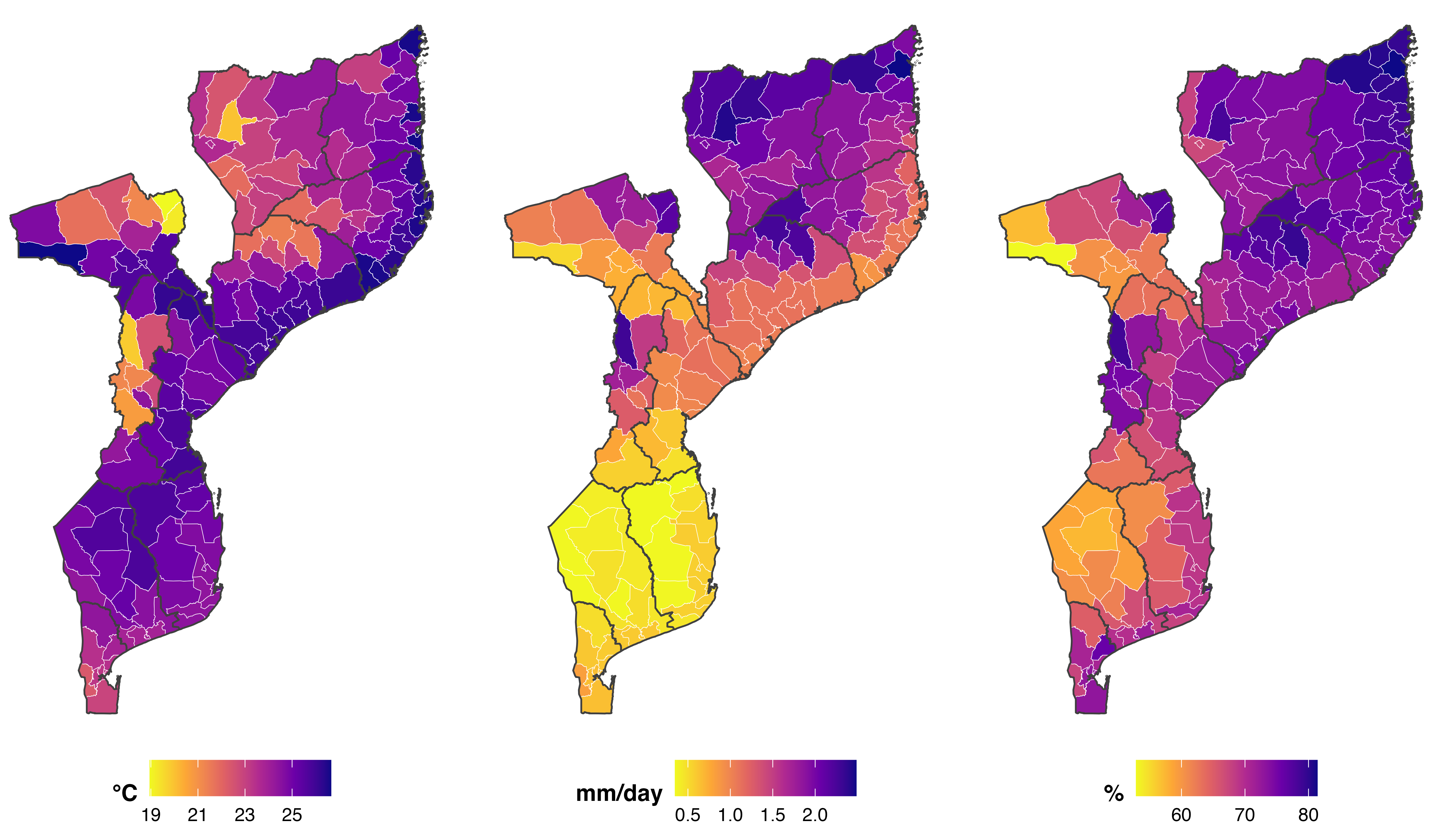}
\caption{District-level averages of daily environmental variables over the period from June 2017 to May 2024. \textbf{Left panel:} average daily temperature (°C). \textbf{Middle panel:} average daily precipitation (mm/day). \textbf{Right panel:} average relative humidity (\%).}
\label{fig:chap4_fig_spatclim}
\end{figure}

\begin{figure}
\centering
\includegraphics[width=0.8\linewidth]{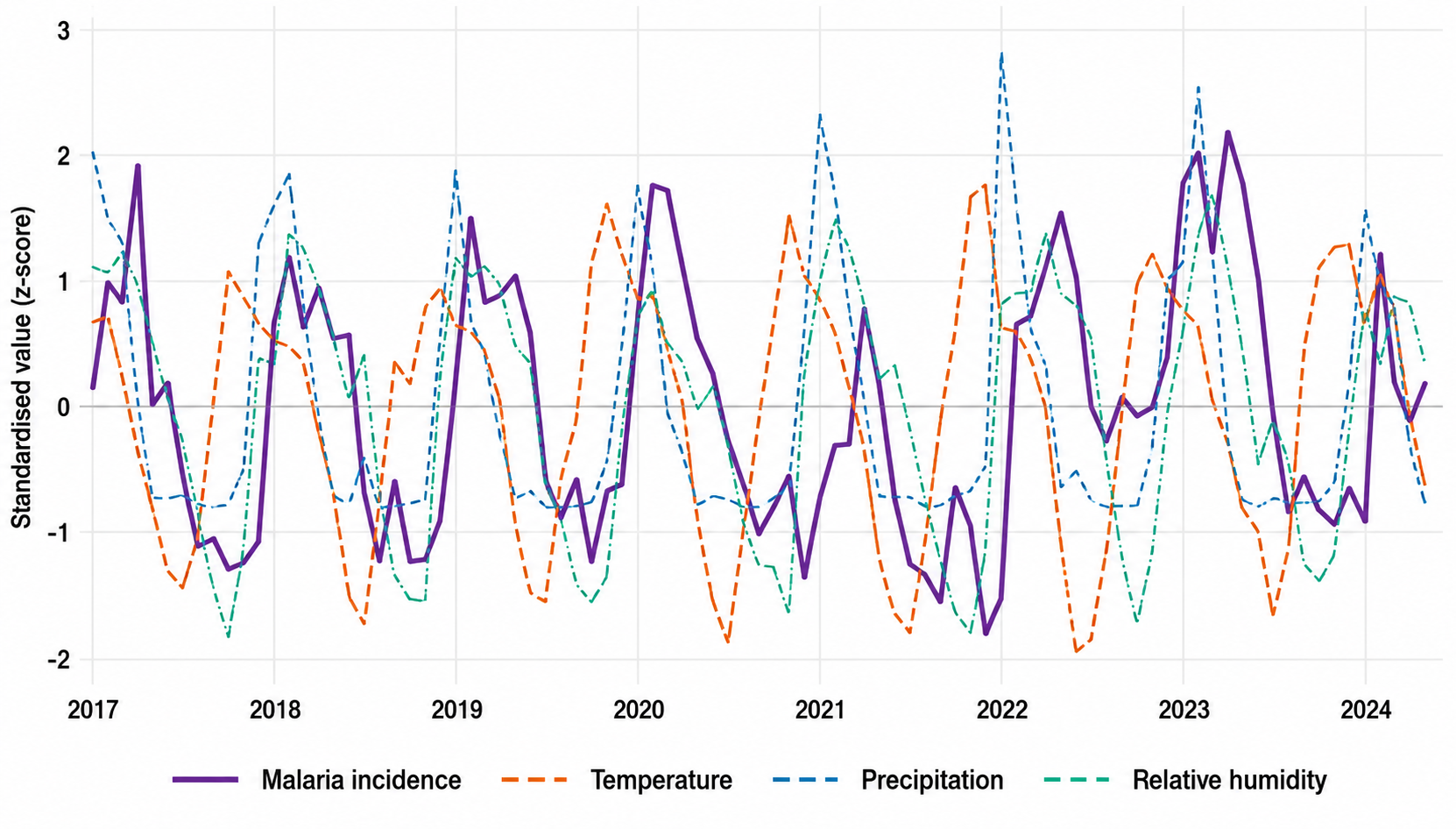}
\caption{Standardized temporal evolution of monthly national malaria incidence and climatic variables between 2017 and 2024. Malaria incidence is shown together with monthly mean temperature, total precipitation, and mean relative humidity.}
\label{fig:chap4_fig_tempclim}
\end{figure}

Figure~\ref{fig:chap4_fig_spatclim} presents the spatial distribution of average temperature, precipitation, and relative humidity at the district level. The maps illustrate substantial spatial heterogeneity in climatic conditions across Mozambique, providing an overview of the geographical variation in the environmental exposures considered in the analysis. In addition to their spatial variation, the climatic variables exhibited distinct temporal patterns throughout the study period. Figure~\ref{fig:chap4_fig_tempclim} presents the standardized temporal evolution of temperature, precipitation, and relative humidity alongside national malaria incidence. Standardization facilitates comparison of the temporal patterns across variables measured on different scales and provides a descriptive overview of how climatic variability coincides with changes in malaria incidence over time. \\

\begin{figure}
\centering
\includegraphics[width=0.4\linewidth]{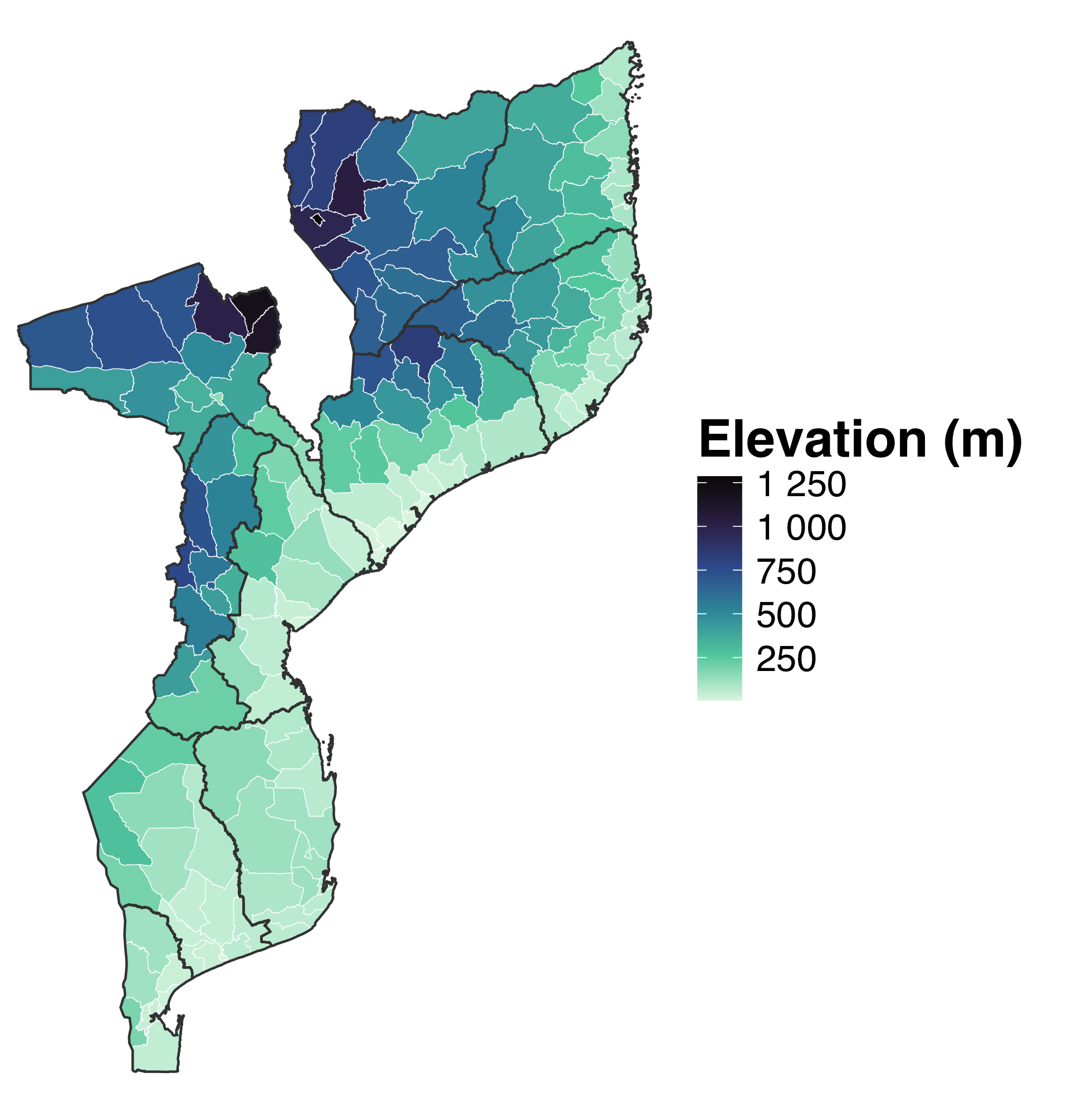}
\caption{Spatial distribution of average district-level elevation in Mozambique.}
\label{fig:ch4_plot_elevation}
\end{figure}

Figure~\ref{fig:ch4_plot_elevation} presents the spatial distribution of average elevation across the districts of Mozambique. Higher elevations are predominantly concentrated in the northern and central-western regions, whereas lower elevations are mainly observed along coastal areas and in southern Mozambique.

\begin{figure}
\centering
\includegraphics[width=0.8\linewidth]{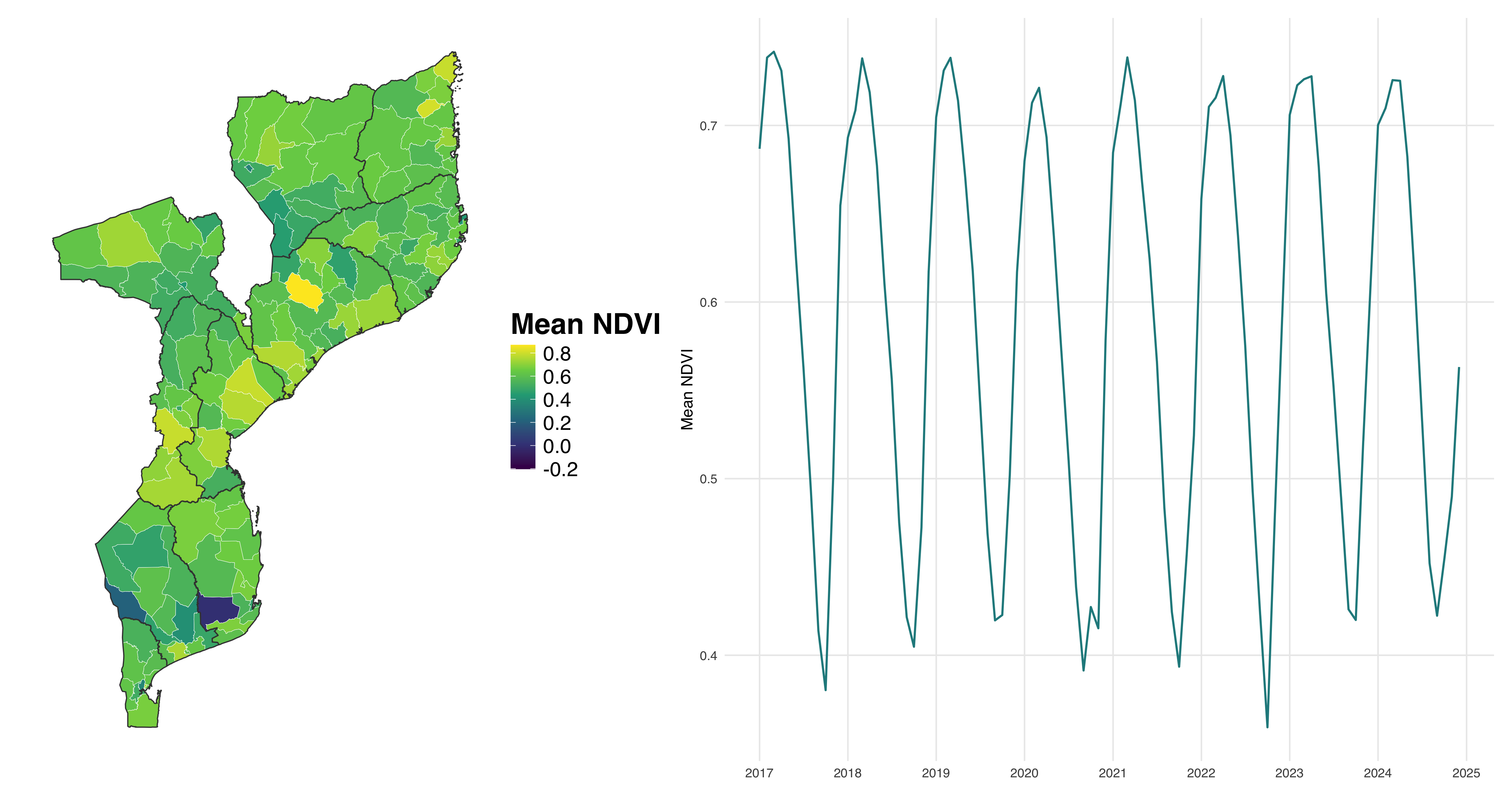}
\caption{Spatial and temporal patterns of the Normalized Difference Vegetation Index (NDVI) in Mozambique during the study period. \textbf{Left panel:} average NDVI for each district. \textbf{Right panel:} national monthly mean NDVI averaged across all districts.}
\label{fig:ch4_plot_ndvi}
\end{figure}

Finally, Figure~\ref{fig:ch4_plot_ndvi} illustrates the spatial and temporal patterns of NDVI during the study period. The spatial distribution indicates generally higher vegetation levels in the northern and central regions and lower values across southern Mozambique. The temporal evolution of the national monthly mean NDVI further illustrates changes in vegetation conditions throughout the study period.

\section{Alternative parameterizations of the random effects}
\label{app:ch4_param_re}

As a sensitivity analysis, three alternative specifications for the latent random effects were considered: two temporal formulations, namely a first-order random walk (RW1) and a temporal Gaussian process (GP), together with the spatio-temporal model proposed by \citet{knorr2000bayesian}. These alternative parameterizations were evaluated to assess the robustness of the proposed framework to different assumptions regarding temporal and spatio-temporal dependence.

\subsection*{First-order random walk}

The structured temporal effect was modeled as a first-order random walk,

\begin{equation}
u_t-u_{t-1}\sim\mathcal{N}(0,\sigma_u^2),
\qquad t=2,\ldots,T,
\end{equation}

where $\sigma_u^2$ controls the variability of successive temporal increments. An identifiability constraint was imposed by fixing $u_1=0$.

\subsection*{Temporal Gaussian process}

Alternatively, temporal dependence was modeled using a Gaussian process,

\begin{equation}
\mathbf{u}\sim
\mathcal{N}\left(\mathbf{0},K\right),
\end{equation}

where $K$ is the $T\times T$ covariance matrix with entries

\begin{equation}
K_{tt'}
=
\sigma^2
\left(
1+\frac{\sqrt{3}|t-t'|}{\rho}
\right)
\exp
\left(
-\frac{\sqrt{3}|t-t'|}{\rho}
\right),
\end{equation}

corresponding to a Matérn covariance function with smoothness parameter fixed at $\nu=1.5$ \citep{matern2013spatial}. Here, $\sigma$ denotes the marginal standard deviation governing the overall variability of the temporal process, while $\rho$ controls the rate at which temporal correlation decays. \\

The temporal domain was rescaled to the interval $[-1,1]$ prior to model fitting to improve numerical stability. Priors were specified as

\begin{equation}
\rho\sim\text{Inverse-Gamma}(5,5),
\qquad
\sigma\sim\text{Half-Normal}(0,1).
\end{equation}

\subsection*{Knorr--Held spatio-temporal model}

Finally, the Bayesian spatio-temporal models proposed by \citet{knorr2000bayesian} were considered,

\begin{equation}
u_{it}
=
\psi_i
+
\zeta_i
+
\gamma_t
+
\upsilon_t
+
\chi_{it},
\end{equation}

where $\psi_i$ denotes the spatially structured random effect modeled using an intrinsic conditional autoregressive (ICAR) prior, $\zeta_i$ is an unstructured spatial random effect assigned an independent Gaussian prior, $\gamma_t$ represents an unstructured temporal random effect, $\upsilon_t$ is a structured temporal random effect modeled as a first-order random walk, and $\chi_{it}$ denotes the spatio-temporal interaction.\\

The interaction term captures residual spatio-temporal variation beyond the additive spatial and temporal main effects, allowing localized departures from the overall spatial and temporal trends. Following \citet{knorr2000bayesian}, four alternative interaction structures were considered according to whether the spatial and temporal components entering the interaction are structured or unstructured. Specifically,

\begin{center}
\begin{tabular}{lll}
\hline
Interaction & Spatial component & Temporal component\\
\hline
Type I & Unstructured & Unstructured\\
Type II & Structured & Unstructured\\
Type III & Unstructured & Structured\\
Type IV & Structured & Structured\\
\hline
\end{tabular}
\end{center}

The four interaction structures were fitted separately and compared as part of the sensitivity analysis.

\section{Sensitivity analysis with different spatio-  temporal random effects parameterization}\label{app:ch4_sensitivity}

As explain in section \ref{methods} a sensitivity analysis was performed to test whether the spatio-temporal random effects parameterization proposed by \cite{MALE} was providing the best fit. Three alternative parameterizations were tested: (M1) a first-order random walk, (M2) a temporal Gaussian process with common marginal variance and temporal range across districts, and (M3) the random effects specification proposed by \cite{knorr2000bayesian}. Note that $\eta^{seas}, \eta^{int}, \eta^{clim}$ were kept in the model comparison as specified in section \ref{methods}.\\

Initially, the alternative Knorr-Held parameterizations were compared to identify the best-performing interaction structure. Table~\ref{tab:chap5_waickh} summarizes the WAIC values for the models with and without temporal disaggregation. Among the four interaction structures, Type III achieved the lowest WAIC under both modeling approaches (with and without temporal disaggregation). These models were selected to be compared with the other parameterizations. 
\begin{table}[ht]
\centering
\begin{tabular}{lcc}
\hline
\textbf{Model specification} & 
\textbf{Without disaggregation }& 
\textbf{With disaggregation} \\\hline
No interaction       & 216788.9 & 216717.6 \\
Interaction Type I   & 216250.1 & 216283.9 \\
Interaction Type II  & 215502.3 & 200024.3 \\
Interaction Type III & 199032.8 & 199982.7 \\
Interaction Type IV  & 200676.6 & 200182.4 \\
\hline
\end{tabular}
\caption{Comparison of the Knorr-Held spatio-temporal model specifications using the WAIC. Results are presented for models fitted without temporal disaggregation, where daily climatic information was first aggregated to the monthly scale, and with the proposed temporal disaggregation framework.
}
\label{tab:chap5_waickh}
\end{table}

Furthermore, Table \ref{tab:chap4_model_comparison} compares the predictive performance of the competing model specifications. M4, the specification proposed by \cite{MALE} achieved the lowest WAIC and the lowest RMSE for both the training and test datasets, indicating the best overall fit and predictive accuracy. \\ 

\begin{table}[h]
\centering
\begin{tabular}{l c cc cc}
\hline
\textbf{Model} &
\textbf{WAIC} &
\multicolumn{2}{c}{\textbf{RMSE}} &
\multicolumn{2}{c}{\textbf{ECP}} \\
\cline{3-6}
 &
 &
\textbf{Train} &
\textbf{Test} &
\textbf{Train} &
\textbf{Test} \\
\hline
M1 & 217460 & 2975.91 & 4063.22 & 0.94 & 0.88 \\
M2 & 216733 & 2942.10 & 5291.37 & 0.94 & 0.96 \\
M3 & 216720 & 2945.88 & 4160.02 & 0.93 & 0.87 \\
M4 & 198422 & 988.23  & 3284.43 & 0.99 & 0.88 \\
\hline
\end{tabular}
\caption{Comparison of the competing model specifications in terms of model fit (WAIC), predictive accuracy (RMSE), and predictive coverage (ECP) in training and test set. M1: first-order random walk temporal effects. M2: temporal Gaussian process random effects. M3: Knorr-Held type III. M4: \cite{MALE}'s proposal.}
\label{tab:chap4_model_comparison}
\end{table}

\section{Exploratory analysis of the between and within components of the environmental variables}
\label{app:ch4_betwith}

\begin{figure}[htbp]
    \centering
    \includegraphics[width=\textwidth]{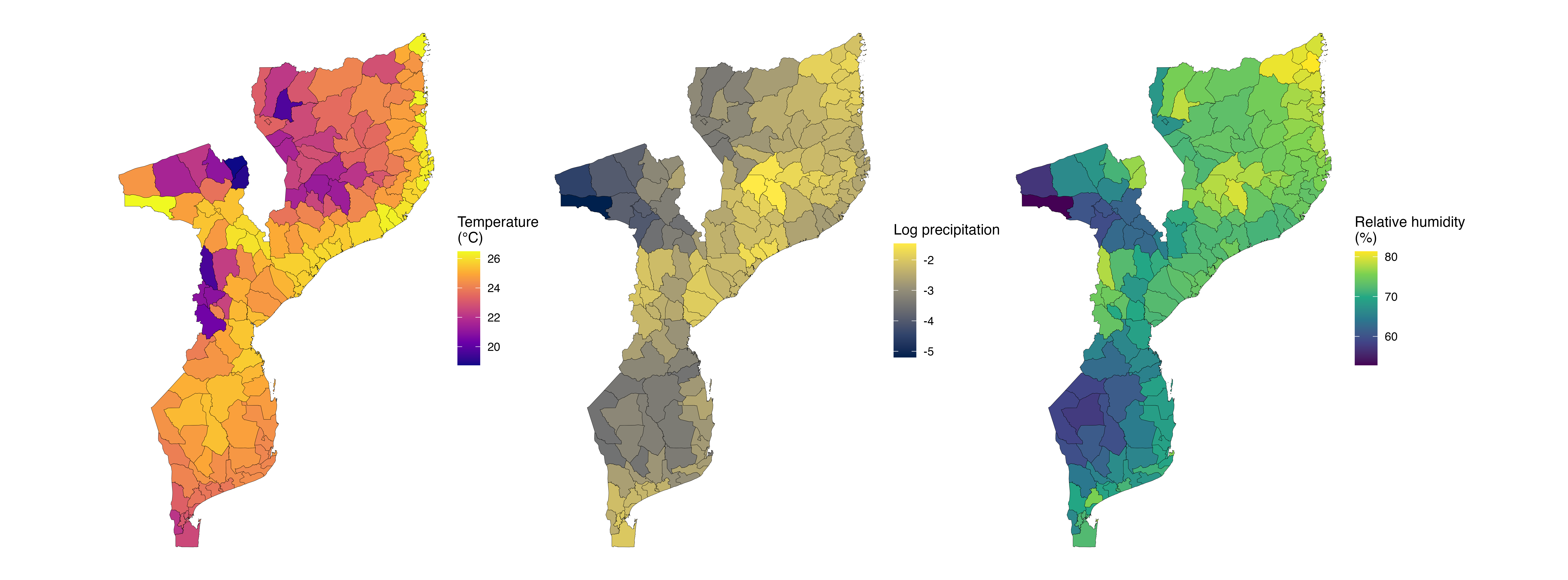}
    \caption{District-specific average climatic conditions over the study period. Panels display the between-district components of (left) daily mean temperature, (center) log-transformed daily precipitation, and (right) daily mean relative humidity.}
    \label{fig:between_maps}
\end{figure}

The between-within decomposition of environmental variables described in section~\ref{chp4.climatic.effect} separates each variable into a between-district component, representing the district-specific average over the study period, and a within-district component, representing the temporal deviation from that average. Figure~\ref{fig:between_maps} illustrates the spatial distribution of the between-district averages for temperature, precipitation, and relative humidity across Mozambique. Clear spatial gradients are observed for all three variables, highlighting the substantial climatic heterogeneity across the study region. \\

\begin{figure}[!b]
    \centering
    \includegraphics[width=\textwidth]{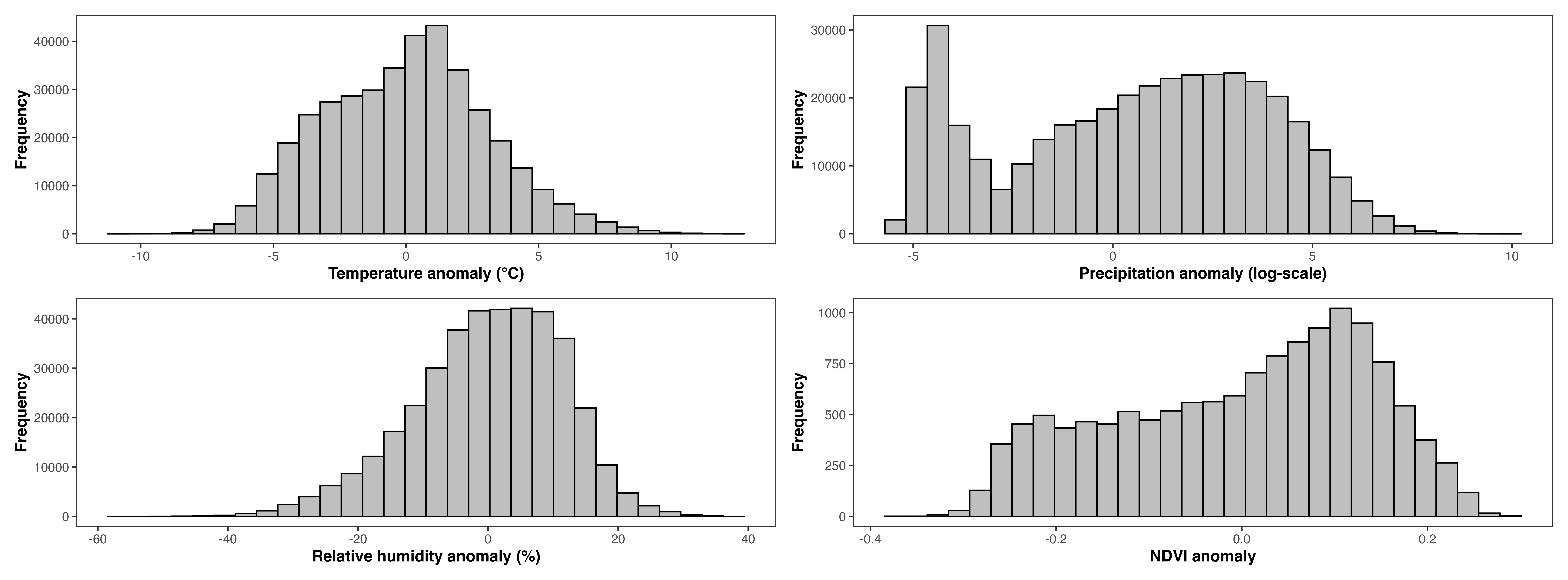}
    \caption{Distribution of the within-district components of the environmental variables.}
    \label{fig:within_histograms}
\end{figure}

The distributions of the within-district components are shown in Figure~\ref{fig:within_histograms}. As expected, the anomaly variables are centered approximately around zero because they represent deviations from each district's long-term mean. These within-district components capture short-term temporal variability while removing the spatial differences represented by the between-district averages.

\section{Assessment of multicollinearity among environmental variables}
\label{app:ch4_coll}
Potential multicollinearity among the environmental variables was assessed in two ways. First, via a quantification of the correlations between the three main climatic variables: daily mean temperature, total daily precipitation, and daily mean relative humidity, while accounting for their within- and between-district decomposition. Second, a sensitivity analysis was conducted by fitting a sequence of nested models with progressively more environmental covariates and evaluated whether the incremental inclusion of these variables altered the estimated exposure–response and lag–response relationships.

\subsection{Correlation exploration}

As described in section \ref{methods}, the between-component of the environmental variables corresponds to the district-specific average over the study period. Pairwise correlations were calculated among the district-level averages of temperature, precipitation, and humidity. The strongest association was observed between mean precipitation and mean humidity ($r=0.83$). Mean temperature was moderately negatively correlated with both mean precipitation ($r=-0.40$) and mean humidity ($r=-0.24$). The relatively strong correlation between precipitation and humidity should be considered when interpreting the corresponding between-district regression coefficients. These coefficients represent conditional associations after adjustment for the remaining environmental variables in the model and should therefore not be interpreted as the total effects of individual climatic factors. \\

The within-components represent temporal deviations from each district's long-term average climatic conditions. To assess correlations at this scale, pairwise correlations among the temperature, precipitation, and humidity anomalies were calculated separately for each district using the complete study period. \\

\begin{figure}
    \centering
    \includegraphics[width=0.8\linewidth]{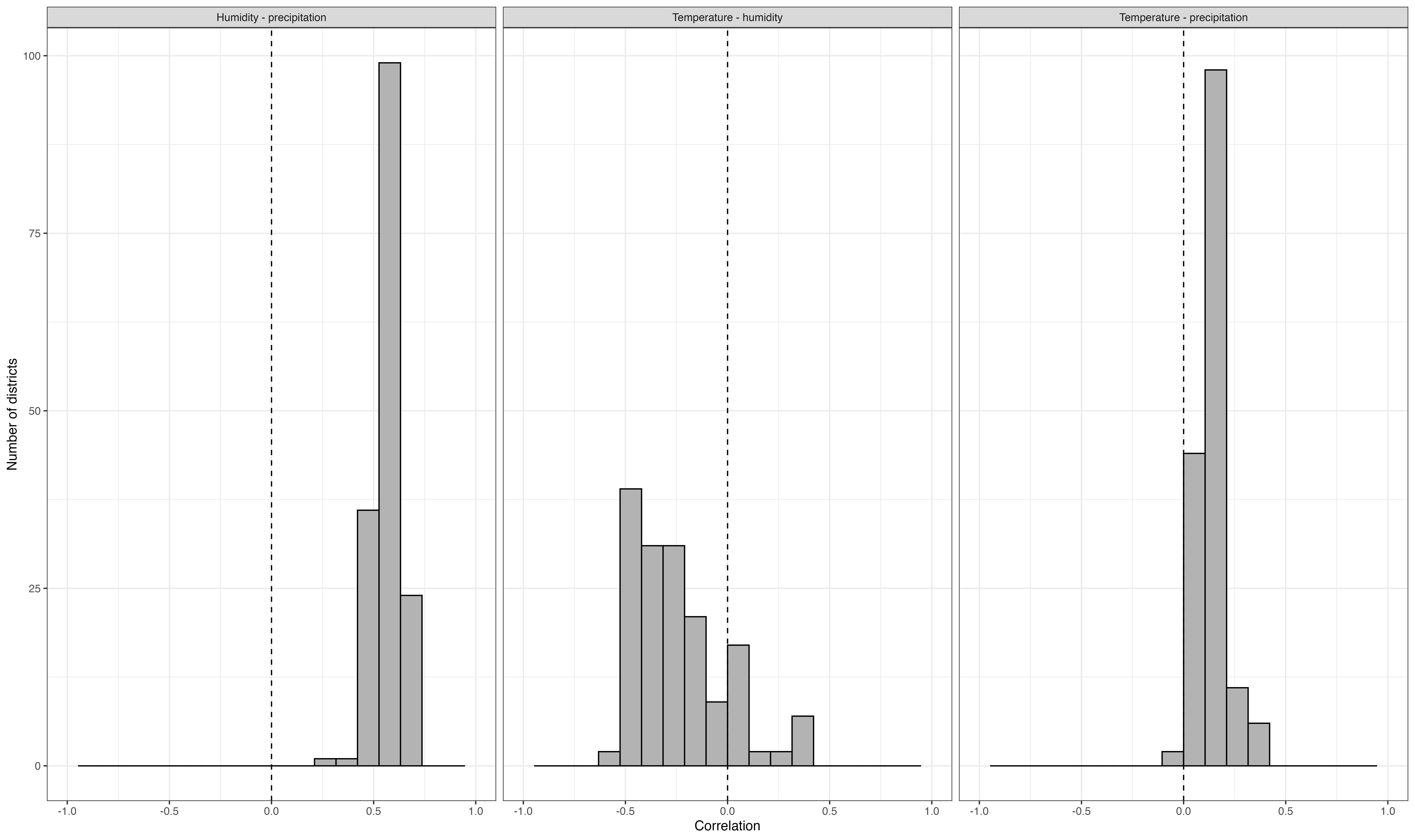}
    \caption{Distribution of the district-specific correlations among the within-components of the enviromental variables.}
    \label{fig:within_climate_correlations}
\end{figure}

Figure~\ref{fig:within_climate_correlations} summarizes the distribution of these district-specific correlations. Compared with the between-district averages, the correlations among the anomaly components were more heterogeneous across districts. Humidity and precipitation anomalies were generally positively correlated, whereas temperature anomalies tended to be negatively correlated with humidity and positively correlated with precipitation. Overall, the within-district anomalies exhibited weaker and more variable correlations than the between-district averages, indicating that although the decomposition reduced shared long-term variation among the climatic variables, some temporal dependence remained.

\subsection{Sensitivity of the distributed lag nonlinear associations}

To assess whether correlations among the environmental variables materially affected the estimated climatic associations, additional models were fitted by introducing the climatic variables sequentially. The exposure-response and lag-response functions obtained from the single-exposure models were compared with those estimated after including additional climatic covariates. Note that all models included the temporal disaggregation approach, the random effects specification proposed by \cite{MALE}, and the intervention covariates. \\

\begin{figure}[htbp]
    \centering
    \includegraphics[width=\textwidth]{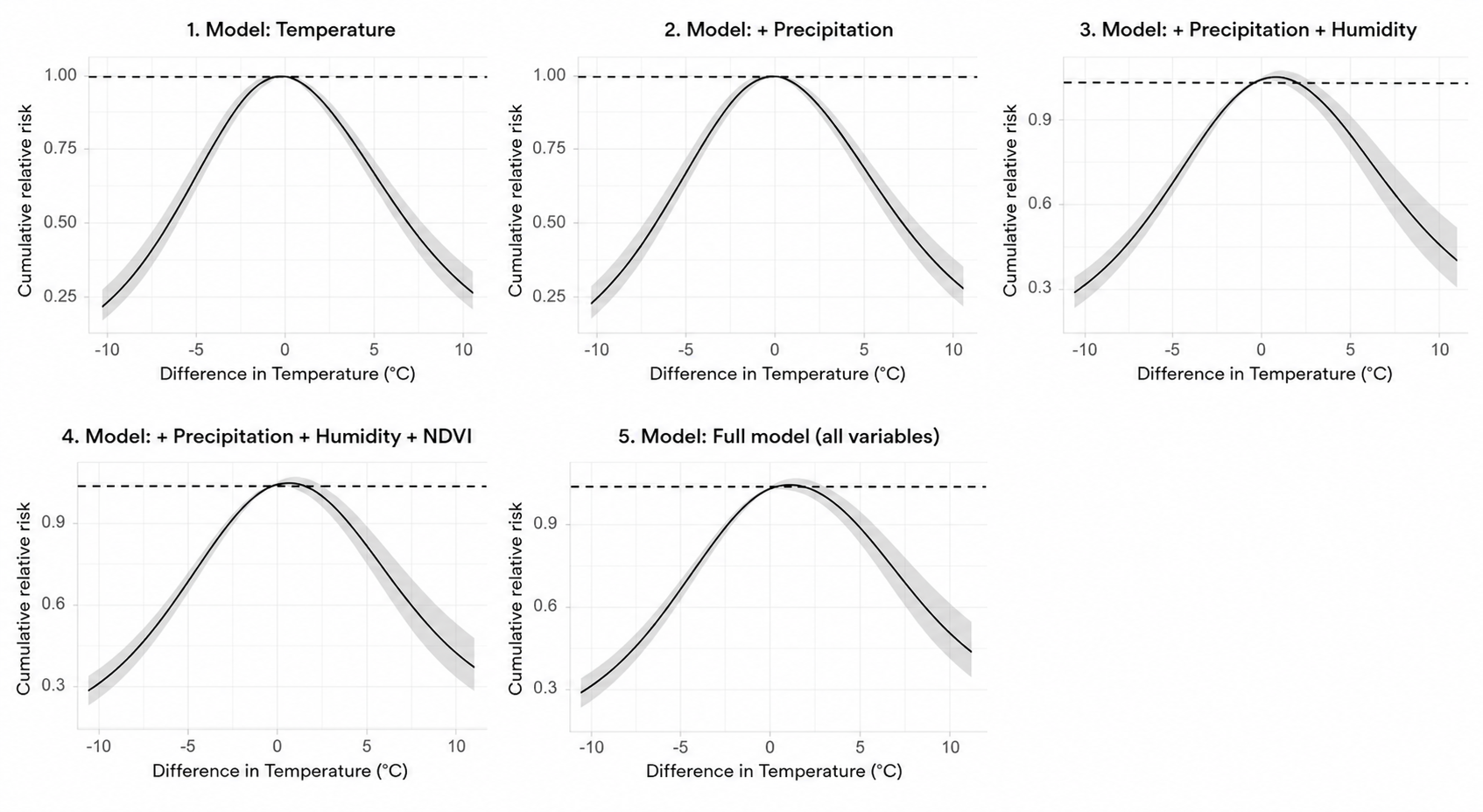}
    \caption{Sensitivity analysis of the cumulative exposure–response relationship for temperature under progressively more complex model specifications, starting with temperature. 95\% credible intervals are included. The horizontal dashed line indicates a relative risk of 1.}
    \label{fig:er_temp_sens}
\end{figure}

\begin{figure}[htbp]
    \centering
    \includegraphics[width=\textwidth]{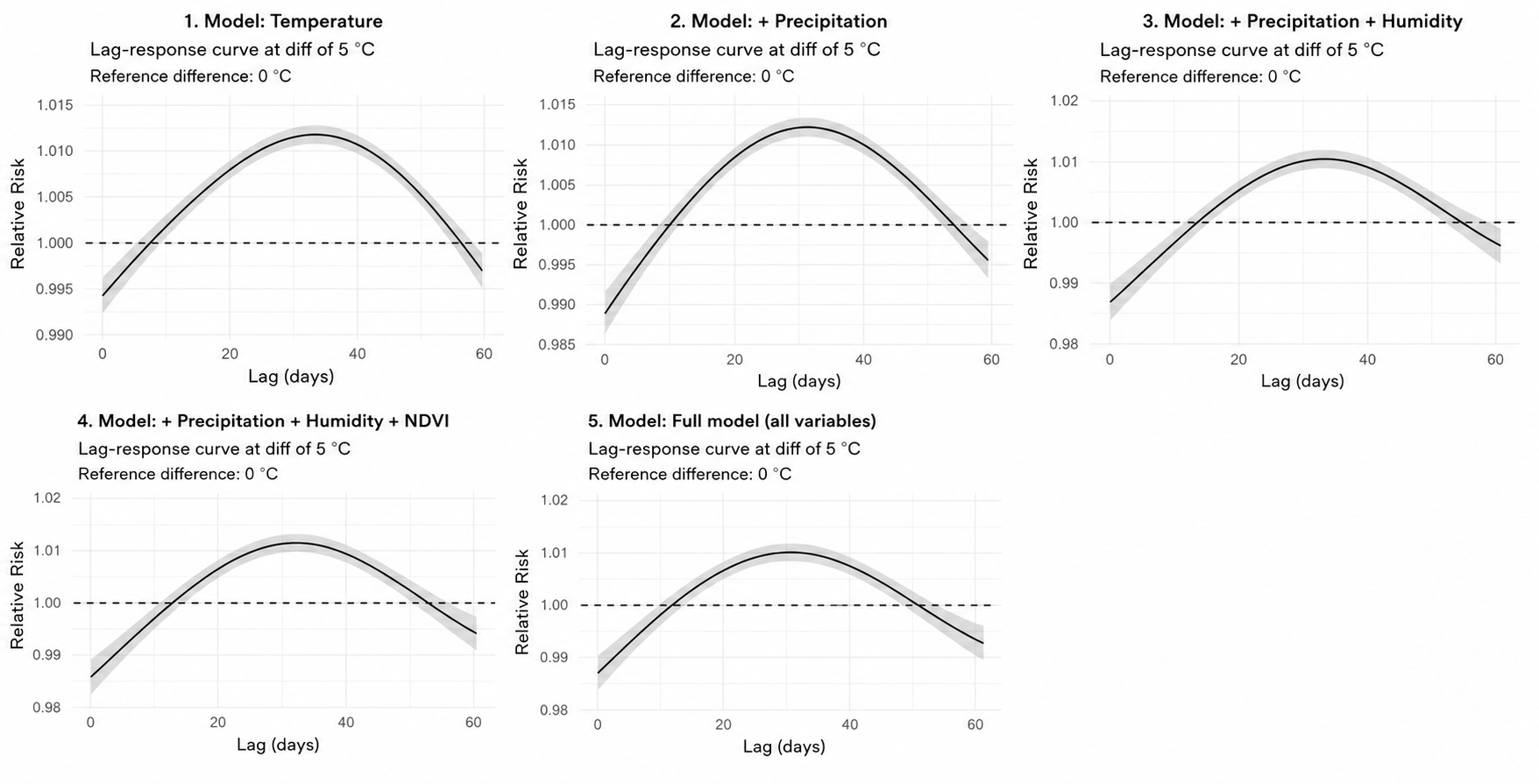}
    \caption{Sensitivity analysis of the lag–response relationship under progressively more complex model specifications, starting with temperature. 95\% credible intervals are included. The horizontal dashed line indicates a relative risk of 1.}
    \label{fig:lr_temp_sens}
\end{figure}

\begin{figure}[htbp]
    \centering
    \includegraphics[width=\textwidth]{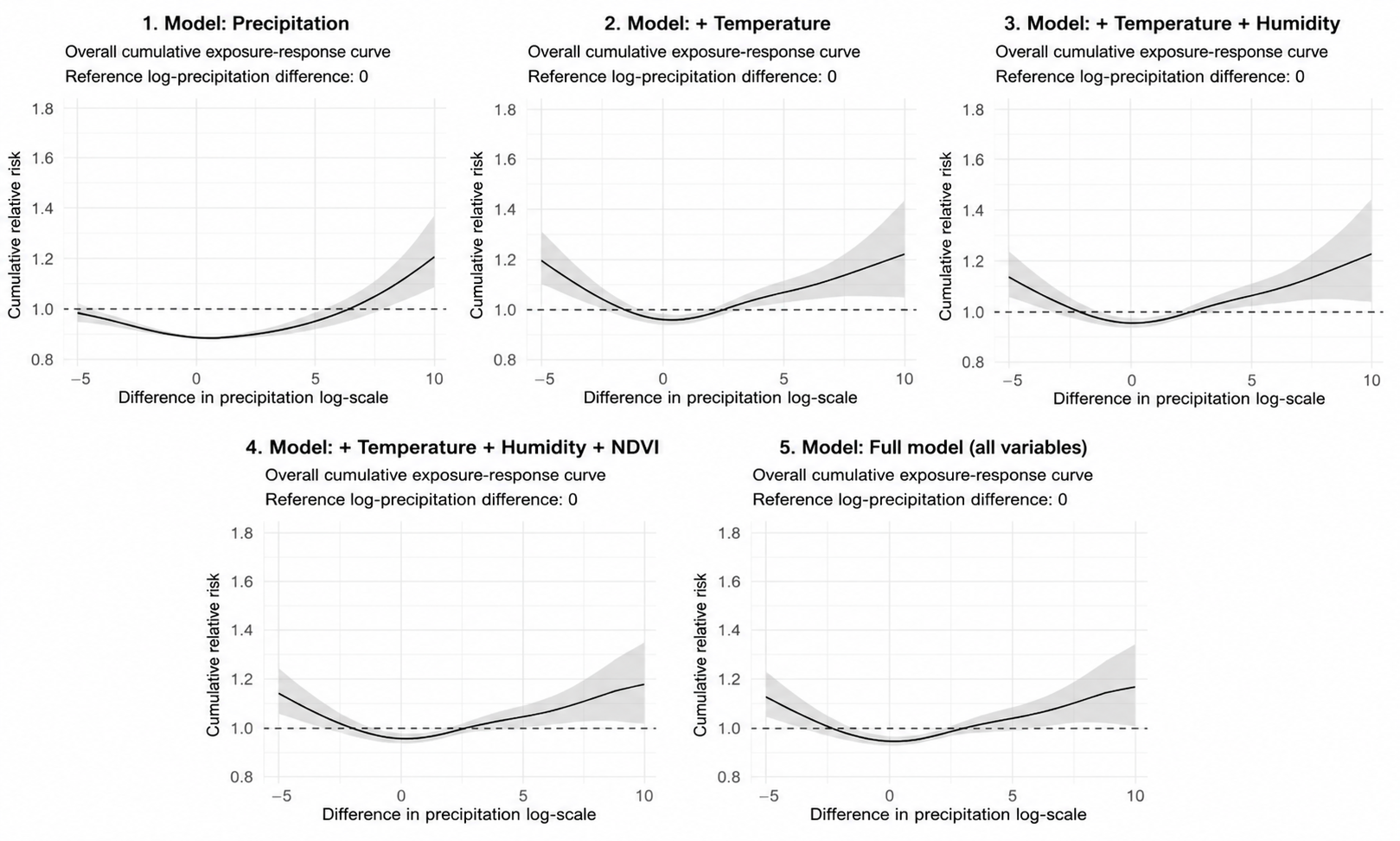}
    \caption{Sensitivity analysis of the cumulative exposure–response relationship for the log-transformed precipitation under progressively more complex model specifications, starting with temperature. 95\% credible intervals are included. The horizontal dashed line indicates a relative risk of 1.}
    \label{fig:er_prec_sens}
\end{figure}

\begin{figure}[htbp]
    \centering
    \includegraphics[width=0.9\textwidth]{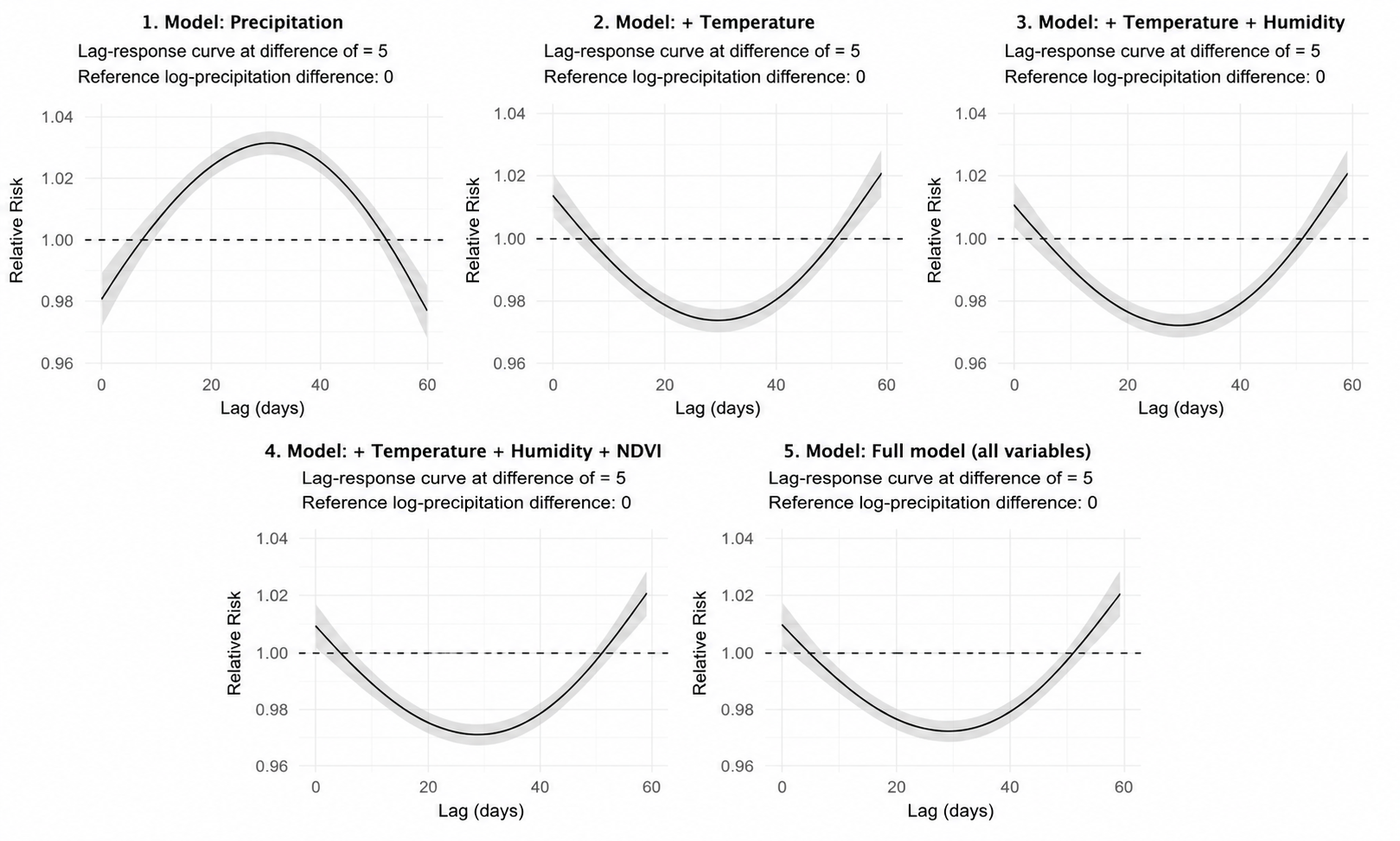}
    \caption{Sensitivity analysis of the lag–response relationship for the log-transformed precipitation under progressively more complex model specifications, starting with temperature. 95\% credible intervals are included. The horizontal dashed line indicates a relative risk of 1.}
    \label{fig:lr_prec_sens}
\end{figure}

\begin{figure}[htbp]
    \centering
    \includegraphics[width=0.8\textwidth]{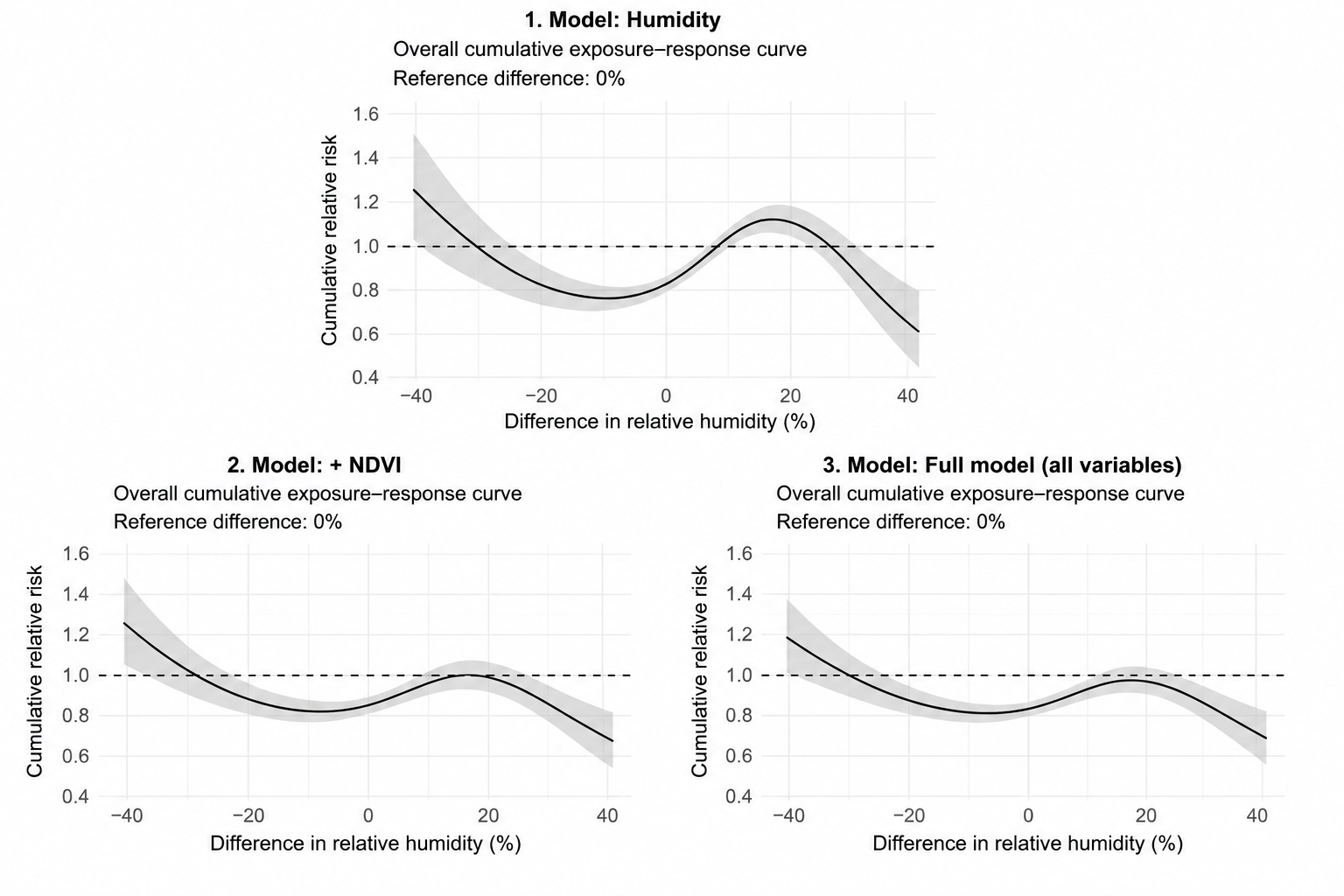}
    \caption{Sensitivity analysis of the cumulative exposure–response relationship for the relative humidity under progressively more complex model specifications, starting with temperature. 95\% credible intervals are included. The horizontal dashed line indicates a relative risk of 1.}
    \label{fig:er_humi_sens}
\end{figure}

\begin{figure}[htbp]
    \centering
    \includegraphics[width=0.8\textwidth]{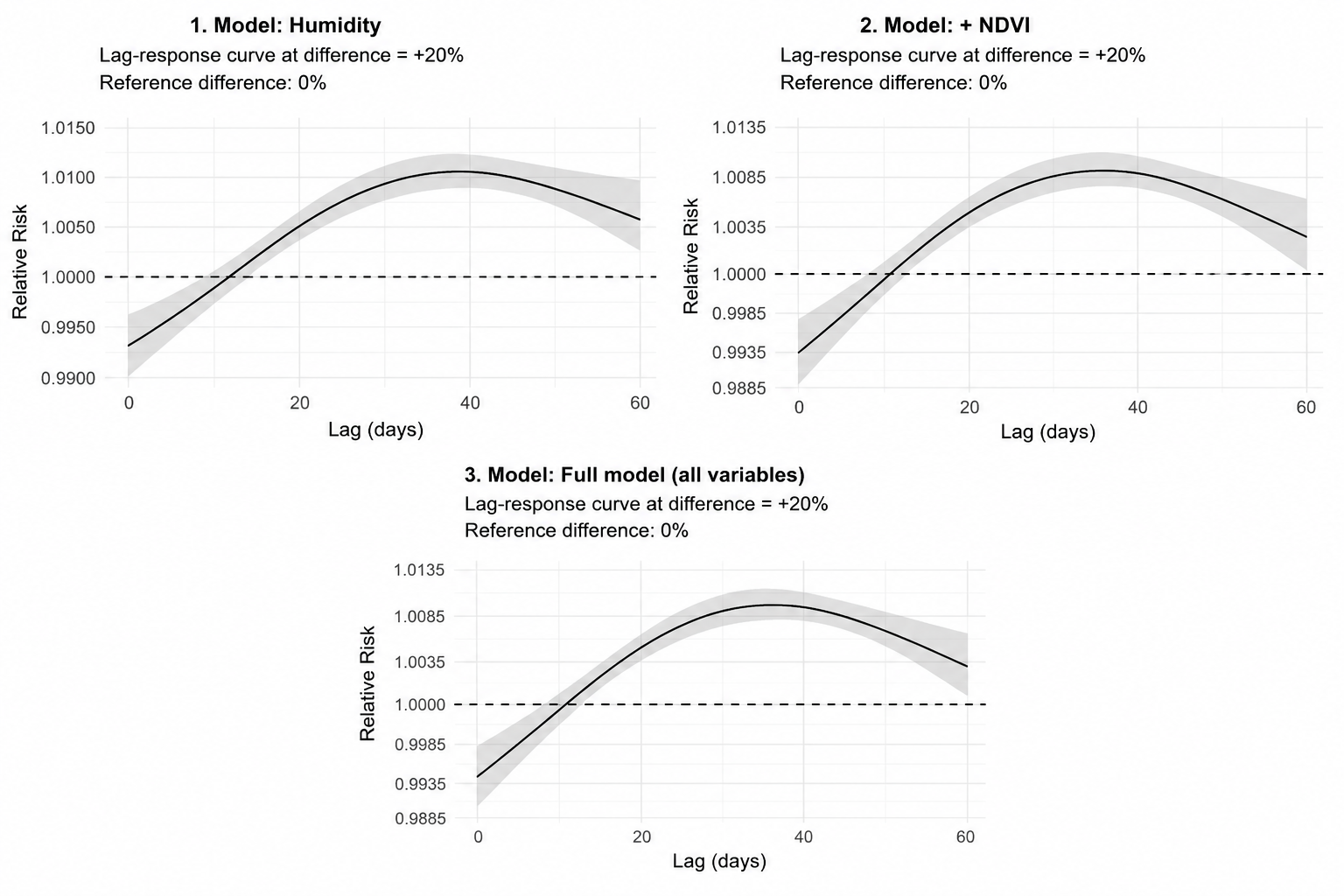}
    \caption{Sensitivity analysis of the lag–response relationship for the relative humidity under progressively more complex model specifications, starting with temperature. The horizontal dashed line indicates a relative risk of 1.}
    \label{fig:lr_humi_sens}
\end{figure}

The estimated functions were generally stable across model specifications (see Figure \ref{fig:er_temp_sens},\ref{fig:lr_temp_sens},\ref{fig:er_prec_sens},\ref{fig:lr_prec_sens},\ref{fig:er_humi_sens},\ref{fig:lr_humi_sens}). The main difference was observed for the precipitation lag-response function when the model containing precipitation alone was compared with the model containing both precipitation and temperature (Figure \ref{fig:lr_prec_sens}). In contrast, the cumulative exposure-response functions and the lag structures of the other climatic variables showed only limited and negligible changes following adjustment for additional exposures. \\ 

Taken together, the analyses show stronger correlation among some between-district climatic averages, particularly precipitation and humidity, but weaker and more spatially heterogeneous correlations among the within-district anomalies. The relative stability of most of the exposure-response and lag-response functions across model specifications provides some reassurance that multicollinearity did not substantially determine the principal findings. However, it remains difficult to isolate the independent contribution of each climatic exposure because temperature, precipitation, and humidity are interconnected components of the broader environmental system.

\end{document}